\documentclass[12ppt,twocolumns]{emulateapj}

\usepackage{amsthm,amssymb,latexsym,graphicx,amscd,amsmath,amsfonts,amscd,enumerate,color}
\usepackage[latin1]{inputenc}

\begin{document}

\shorttitle{Collapse of Turbulent Cores}
\title{Collapse of Turbulent Cores and Reconnection Diffusion}
\author{M. R. M. Le\~ao\altaffilmark{1},  E. M. de Gouveia Dal Pino\altaffilmark{1}, \\ R. Santos-Lima\altaffilmark{1}, and A. Lazarian\altaffilmark{2}  }
\altaffiltext{1}{\bf Instituto de Astronomia, Geof\'{i}sica e Ci\^encias Atmosf\'ericas, Universidade de S\~ao Paulo, R. do Mat\~ao, 1226, S\~ao Paulo, SP 05508-090, Brazil}
\altaffiltext{2}{\bf Department of Astronomy, University of Wisconsin, Madison, WI 53706, USA}
\email{mrmleao@astro.iag.usp.br\\dalpino@{\bf astro.iag.usp.br} \\rlima@{\bf astro.iag.usp.br}\\{\bf alazarian@facstaff.wisc.edu}}

\begin{abstract}
For a molecular cloud clump to form stars some transport of magnetic flux is required {\bf from the} denser, {\bf internal} regions to the outer regions, otherwise this can prevent the gravitational collapse. Fast magnetic reconnection which takes place in the presence of turbulence can induce a process of reconnection diffusion that has been elaborated in  earlier theoretical  work.  We have named this process turbulent reconnection diffusion, or simply RD. This paper continues our numerical study of this process and its implications. In particular, extending our studies of reconnection diffusion in cylindrical clouds we consider more realistic clouds with spherical gravitational potentials (from embedded stars) and also account for the effects of the gas self-gravity. We demonstrate that within our setup reconnection diffusion is efficient. We have also identified the conditions under which reconnection diffusion becomes strong enough to make an initially subcritical cloud clump supercritical and induce its collapse. Our results indicate that the formation of a supercritical core is regulated by a complex interplay  between gravity, self-gravity, the magnetic field strength and nearly transonic and trans-Alfv\'enic turbulence and therefore, is very sensitive to the initial conditions of the system. In particular, self-gravity  helps reconnection diffusion and, as a result, the magnetic field decoupling from the collapsing gas becomes more efficient compared to the case of an external gravitational field. Our simulations confirm that reconnection diffusion {\bf can transport magnetic flux from the core of collapsing} clumps {\bf to the envelope}, but only a few of them become nearly critical or supercritical, sub-Alfv\'enic cores, which is consistent with the observations. Besides, we have found that the supercritical cores built up in our simulations develop a predominantly helical magnetic field geometry which is also consistent with recent observations. Finally, we have also evaluated the effective values of the turbulent reconnection diffusion coefficient in our simulations and found that they are much larger than the  numerical diffusion, especially for initially trans-Alfv\'enic clouds, thus ensuring that the detected {\bf magnetic flux} removal {\bf is} due {\bf to the} action {\bf of the turbulent reconnection} diffusion rather than to numerical diffusivity.
\end{abstract}



\section{Introduction}
\label{intro}

It is well {\bf known} that star formation occurs in dense globule-like regions inside giant molecular clouds \citep{Blitz1993,Williams2000}. This is due to their high densities and low temperatures which help the gravitational force to overcome the  outward forces which act to prevent collapse. However, in spite  of all observational and theoretical efforts, it is not yet fully clear how these globules form stars.

Two key ingredients are  present in the clouds: magnetic fields and turbulence. For years it has been believed that the magnetic pressure gradient is an important agent working against the gravitational collapse of the gas and making star formation inefficient (see Mestel \& Spitzer 1956, Mestel 1965, Mouschovias 1991). {\bf The mass-to-magnetic flux ratio, $M/\Phi \simeq N/B$, where M is the cloud mass, N is the column density, and B is the magnetic field, quantifies the stability of a magnetized cloud defining how strong a static magnetic field must be to support the cloud against gravitational collapse} (e.g., Nakano \& Nakamura 1978, Crutcher 1999, 2005a, Crutcher \& Troland 2007, Heiles \& Crutcher 2005, Falgarone et al. 2008, Crutcher et al. 2010b). Considering an initially spherical cloud, the critical value of this ratio implies $B_{cr} \simeq 2.57\times (N/10^{21}$ cm$^{-2})$ $\mu$G, for hydrogen gas. {\bf Zeeman measurements} suggest {\bf that the mass-to-magnetic flux ratios in molecular cloud cores are} around 2.0 times larger than the critical value ($\mu$) for which gravity and magnetic support balance each other \citep{crutcher2008}.

Another essential element of the star formation picture is turbulence. Indeed, turbulence is known to be ubiquitous in the interstellar medium \footnote{Turbulence in interstellar medium is not only theoretically expected due to the high Reynolds numbers of the fluid (see e.g., {\bf Lazarian et al. 2012 and ref. therein}), but also, observed. In fact, turbulence spectra are measured in the ionized component of the medium {\bf (see Armstrong et al. 1995, Chepurnov \& Lazarian 2009)}, neutral hydrogen (see Lazarian \& Pogosyan 2000, Stanimirovic \& Lazarian 2001, Chepurnov et al. 2010) {\bf and CO lines} (see Padoan et al. 2006, 2009).} and it is currently widely accepted that molecular clouds are part of the big cascade. Turbulence is believed to rule the structuring and fragmentation of the molecular clouds in all scales, {\bf and can} also {\bf have a dominant role during the different phases of star formation \citep{MacLow2004,mckee2007}}. The observation of broad line widths ranging from a few to more than 10 times the sound speed indicate that the  turbulent motions are mainly supersonic to transonic in the molecular clouds. This means that turbulence is roughly in equipartition with the magnetic fields in the interstellar medium and for this reason it is believed to be trans-Alfv\'enic (Elmegreen \& Scalo 2004; Heiles \& Troland 2005). In fact, the inferred ratio between turbulent and magnetic energy for the cold neutral interstellar gas is roughly $1.3 < E_{turb}/E_{mag} < 1.9$ (Heiles \& Troland 2005). The formation of structures appears to be related to compression caused by the compressional motions associated with the turbulent cascade. The relation between compressible and incompressible parts of the cascade depends on the sonic and Alfv\'en Machs of the turbulent flow (Cho \& Lazarian 2002, 2003). Even if the turbulence is globally strong enough for supporting the cloud against gravity {\bf \citep{Klessen2000, MacLow2004, Vazquez2005}}, compressible motions could also cause the collapse at small scales, by generating high density regions in both magnetized \citep{Heitsch2001,Nakamura2005,Melioli2006a, Leao2009} and {\bf non-magnetized} medium {\bf \citep{Klessen2000,Elmegreen2004}}. 

Turbulent stirring happens at different scales, both due to the processes acting inside and outside molecular clouds. There are several candidate mechanisms for  injection of turbulent energy inside a cloud. Internal sources include  feedback from low-mass (e.g., {\bf de Gouveia Dal Pino} 1999; Raga {\bf et al.} 2002) and massive stars. The latter in particular, can induce turbulence through intense ionizing radiation, winds, and supernovae explosions (SNe) \citep{McCray1979, Melioli2006a, Melioli2006b, Leao2009}. External candidates also include {\bf SNe shocks} \citep{Wada2001,Elmegreen2004,Melioli2006a,Melioli2006b, Leao2009}, magneto-rotational instability (Fleck 1981; Sellwood and Balbus 1999), and the {\bf galactic spiral} structure {\bf \citep{Roberts1969,Bonnell2006}}. All {\bf these processes seem to have} power enough to explain the structuring and {\bf kinematics of the ISM, and} could {\bf generate the observed dispersion} of the gas velocities \citep{Kornreich2000}. Other sources of turbulence include proto-stellar jets and winds (e.g. de Gouveia Dal Pino 1999; Raga et al. 2002), expansion of HII regions and establishing their relative importance requires further studies \citep{Joung2006, Ballesteros2006, MacLow2009}.

For decades magnetic fields and turbulence were considered separate elements of the star formation picture. It has been thought that it is magnetic field that makes star formation inefficient. Assuming that in highly conducting media magnetic flux should be perfectly frozen in, the researchers were searching for the way of removing this flux discussing different stages of star formation and faced many tough problems, from magnetic breaking catastrophe that we discuss below to a famous problem of young stars being magnetized orders of magnitude higher than is observed. Indeed, if all the magnetic flux passing through the clouds {\bf was advected with the collapsing  material,} {\bf then the magnetic field in a protostar would be} the order of a pulsar, i.e. much higher than it is the {\bf observed in T-Tauri stars. This is } the well known textbook "magnetic flux problem" {\bf (see Galli et al. 2006, Johns-Krull 2007,  Santos-Lima et al.2012, 2013) which illustrates the difficulties of treating magnetic} fields in star formation process.

Traditionally, {\bf in order to deal with the problem of magnetic flux diffusion both in the partially ionized interstellar medium and in molecular clouds,} {\bf researchers have relied on the ambipolar diffusion (AD)} mechanism (e.g., Mestel \& Spitzer 1956; Spitzer 1968; Nakano \& Tademaru 1972; Mouschovias 1976, 1977, 1979; Nakano \& Nakamura 1978; Shu 1983; Lizano \& Shu 1989; Fiedler \& Mouschovias 1992, 1993; Li et al. 2008; Fatuzzo \& Adams 2002; Zweibel 2002).
\textbf{ In principle, during the collapse of low ionization regions, magnetic flux is redistributed by AD as the result of the  slippage between} the ions and neutral atoms. New theoretical studies, however, have evidenced that the process has problems to explain observational data. For instance, {\bf the accretion phase in low-mass star formation was explored by} several authors (Shu et al. 2006, \citealt{Krasnopolsky2010,Krasnopolsky2011}) {\bf who concluded that 
the AD would work only under special circumstances, i.e.,  considering specific dust grain sizes  (Li et al. 2011) in order to produce the required effective diffusivity around three orders of magnitude larger than the Ohmic diffusivity.} 
Besides, recent results from observations of cloud cores have also challenged predictions from AD (\citealt{crutcher2010b}, see also Section \ref{observations}). In other words,  it is evident that AD faces difficulties solving the magnetic flux transport problem in collapsing flows.

The underlying assumption to all the studies above is that magnetic field is well frozen in within highly conductive fluids which seems to be the fact established by Hans Alfv\'en (1942) famous theorem. There is, however, a problem with this sort of reasoning. Solar magnetic fields embedded in highly conducting gas show exhibit process of magnetic reconnection, which evidently violates  the flux freezing. Therefore, it is natural to argue that magnetic reconnection may be also important for other processes, including star formation. The big impediment for such an approach was the poor understanding of magnetic reconnection. This process stayed enigmatic in spite of intensive research efforts. In most cases researchers attempted to find out some special conditions pertinent to the Sun, i.e. its atmosphere having collisionless plasmas that would make magnetic reconnection possible (see Shay et al. 1998). Such solutions would not be applicable to the dense collisional media of molecular clouds.

Fortunately, a universal mechanism of fast reconnection has been identified in Lazarian \& Vishniac (1999, henceforth LV99). Their process appeals to magnetic field wandering induced by turbulence and does not depend on the collisional state of the flow, or its resistivity, temperature, etc. Being universally applicable to turbulent magnetized media, it challenges the standard Alfv\'en (1942) theorem and, as it was shown in Eyink, Lazarian \& Vishniac (2011) it also means that the Alfv\'en theorem is only applicable to laminar flows.\footnote{It should be remarked that the frozen in condition is statistically preserved even in turbulent flows (Eink 2011).} In view of this successful challenge of the existing flux freezing paradigm, which also included testing of the analytical predictions of the LV99 reconnection rates in Kowal et al. (2009, 2012), it became natural to consider to what extent the old star formation paradigm based on flux freezing holds for realistic turbulent media. The first ideas challenging the traditional flux freezing and suggesting that magnetic reconnection can remove magnetic field from molecular clouds and accretion disks were presented in Lazarian (2005) (see also Lazarian \& Vishniac 2009 for a discussion of magnetic field removal from an accretion disk via reconnection). However, it was only after numerical demonstration of the efficiency of the processes in simulated molecular clouds (Santos-Lima et al. 2010) and accretion disk formation (Santos-Lima et al. 2012, 2013) that the process that was termed "reconnection diffusion" (see Lazarian et al. 2010; Santos-Lima et al. 2010) became focus of intensive debates. The theoretical foundations of the reconnection diffusion are formulated in Lazarian (2011), and Lazarian et al. (2012a) (see also Gouveia Dal Pino et al. 2012; 2013). In the present work, we provide an additional numerical evidence that reconnection diffusion is important for molecular clouds. In what follows we occasionally abbreviate reconnection diffusion as {\it RD}, in analogy with the accepted abbreviation for the ambipolar diffusion as {\it AD}. Reconnection diffusion implies magnetic field diffusion in turbulent conducting fluids, but this is implied and not a limitation, as turbulence presents both the generic case for the interstellar medium and the necessary requirement for the LV99 fast reconnection (and consequent violation of the Alfv\'en theorem).

Without going into details of the reconnection diffusion process (see our short discussion in the Appendix  and references therein), we would like to stress {\bf that the problem faced in star formation is of magnetic $diffusion$, not of $dissipation$ of magnetic flux, in which, indeed, ordinary resistivity is necessary.} The ideal non-diffusive MHD condition holds in the absence of turbulence and magnetic reconnection, but {\bf when the latter are present, there are changes in the magnetic field topology}. This is the problem that is being handled by magnetic reconnection in turbulent media. 

As we mentioned earlier, the first numerical \textbf{test} of turbulent reconnection diffusion was performed by Santos-Lima et al. (2010) {\bf who examined the transport of magnetic field flux in  turbulent molecular clouds embedded in a central gravitational potential.}
\textbf{These high-resolution three-dimensional (3D) MHD simulations have evidenced that  
in the presence of turbulence there is a decrease  of the magnetic flux-to-mass ratio with increasing density at the center of the gravitational well caused by the transport of the  magnetic flux to the outskirts of the cloud by RD.} 
 \textbf{The comparison of these results with systems without turbulence revealed no change in the magnetic-flux-to-mass ratio, as one should expect for ideal MHD systems. This was a further evidence that in the cases with turbulence the transport is due to RD.} 
\textbf{ This effect was observed in gas and magnetic field distributions  starting in equilibrium or in dynamically unstable (gravitationally collapsing) configurations, thus indicating that the process of turbulent RD can be applied, in principle, to both quasi-static subcritical clouds or to collapsing supercritical ones.}
These results were also found to be insensitive to the numerical resolution. Tests made with resolutions of $128^3$, and $512^3$ gave essentially the same results as those of $256^3$ thus confirming the robustness of the results and the method. In fact, the effective RD evaluated from the simulations was found to be larger than the numerical (and Ohmic) diffusion especially when turbulence is initially trans-Alfvénic (see Section \ref{ohmic} for more details).

The study above, however, assumed for simplicity clouds with cylindrical gravitational fields. We here perform 3D MHD high resolution numerical simulations considering  more realistic initially subcritical clouds with central spherical gravitational potentials (due to embedded stars) and also account for the effects of self-gravity. Our results essentially confirm the results of the previous study above, but also reveal the conditions under which reconnection  diffusion is efficient enough to make an initially subcritical cloud to become supercritical and collapse. Preliminary results of this study were presented in de Gouveia Dal Pino et al. (2012).

In Section \ref{Setup}, we describe the setup and the numerical methodology that we employed for performing 3D MHD simulations of molecular cloud cores or clumps. In Section \ref{Results}, we present the results of the diffusion of magnetic flux by turbulence considering molecular clouds  with self-gravity and a central spherical gravitational potential. In Section \ref{Discussions}, we discuss our results and compare them with previous works and observations, and in Section \ref{Conclusions} draw our conclusions.

\section{Physical Setup and Numerical Methodology}
\label{Setup}

The astrophysical environment we want to investigate in this work consists of an initially $subcritical$ molecular cloud clump with a small group of embedded stars sustained by magnetic field and turbulence. Our goal is to examine, by means of three-dimensional MHD numerical simulations, the conditions under which the transport of magnetic flux by reconnection will allow the contraction of the self-gravitating gas to form a supercritical core.  We also consider a few models without including self-gravity in order to compare with the previous study  of Santos-Lima et al. (2010). In most of our numerical experiments, the system starts already out-of-equilibrium between gravity and the other forces. {\bf A gas clump immersed in a initially homogeneous magnetic field undergoes a fast contraction (for a period of the order of the free-fall time) due to the presence of a spherical central gravitational potential that mimics a small group of embedded stars, after which the evolution goes on more smoothly due to  the injected turbulence.} 
As an example, Figure 1 shows the logarithmic density map of the central slice of model N2b (with no \textbf{turbulence growth} yet) right  after an initial fast contraction ($\sim 1.1$ Myrs). The magnetic field configuration is superimposed to the density map.

We also present one test with the clump starting in magneto-hydrostatic equilibrium for comparison with the other models. (In this case, we actually started with a uniform density cloud, let it relax to the equilibrium and then injected turbulence). 

{\bf The simulation of each system is performed inside a cubic domain with periodic boundaries.} Since a typical giant molecular cloud has several clumps the use of periodic boundaries to model one of these clumps is appropriate. {\bf For simplicity, an isothermal equation of state with a single temperature for the whole system is employed,} which means that it radiates quite efficiently.
 
 \begin{figure} [!htb]
  \begin{center}
		\includegraphics[width=0.6\linewidth]{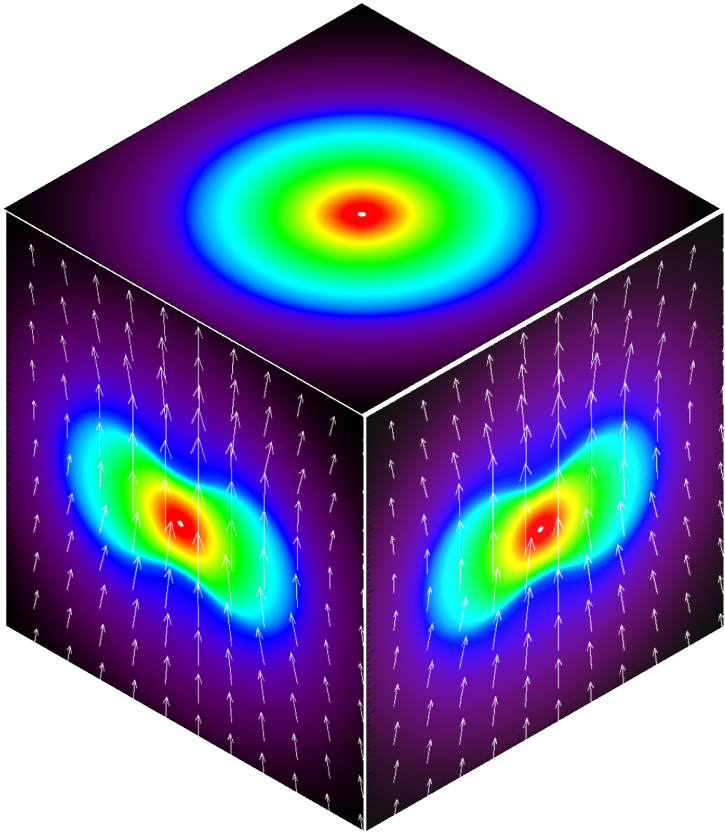}
  \caption{Example of a cloud setup in the computational domain (with no turbulence \textbf{developed yet}) right after an initial fast contraction (at $\sim 1.1$ Myrs). Logarithmic density map and magnetic field lines of the central slice (projected onto the walls of the cubic computational domain) as in model N2b of Table 1 with $M_{pot}=40.7$M$_{\odot}$, initial density $n_0 =90 $ cm$^{-3}$, and $\beta = 3.0$.}
 \end{center}
\label{fig:inicial}
\end{figure} 

We chose characteristic values for the physical parameters of our models based on observational data (see, e.g. Mac Low \& Klessen 2004)\footnote{We note that in this study since we are considering self-gravity, the results are not scale independent, therefore, a physical parameter space must be considered.}. The computational domain has sides $L=3.25$pc. For a grid resolution with $256^3$, the corresponding cell length is $\approx 0.013$pc$ \sim 2600$ AU. As remarked in Section \ref{intro}, in the previous numerical study of 3D MHD reconnection diffusion in cylindrical clouds by Santos-Lima et al. (2010), tests with different resolutions between $128^3$, and $512^3$ were performed and were found to produce essentially the same results, therefore in most of the simulated models here we considered a grid resolution of $256^3$ (see however Fig. \ref{fig:resol} below). 

The isothermal sound speed is $c_s = 3 \times 10^{4}$ cm/s, which corresponds to a temperature $T=10.9$ K (for a mean molecular weight $\mu=1.0$). In order to cover a parameter space as much appropriate as possible to the observed conditions we considered an initial uniform gas density, $\rho_0$, spanning between  $1.67 \times 10^{-23}$ g cm$^{-3}$ (or $n_0 = 10$ cm$^{-3}$) and  $1.67 \times 10^{-22}$ g cm$^{-3}$ (or $n_0 = 100$ cm$^{-3}$). We have also considered values of the initial ratio between the thermal and the magnetic pressures, $\beta =c_s^2 \rho /(B^2/8\pi)=$ $3.0, 1.0$ and $0.3$. This implies initial Alfv\'en velocities $v_{A} \approx 2.4\times 10^4$cm/s, $ 4.2\times 10^4$cm/s, and $ 7.8\times 10^4$cm/s, respectively, and initial uniform magnetic field  intensities between $B_{0}=0.35$ $\mu G$ and $3.4$ $\mu G$. 

It should be noticed that, although an average value of $\sim$6 $\mu G$ is usually considered, it has been   claimed (Heiles \& Troland 2005) that there are  substantial variations in the magnetic field value in the cold neutral medium (CNM). Besides, with  the  choice of the initial conditions above we have tried to cover a parameter space as much suitable as possible to the observed conditions while  avoiding  straightforward solutions such as immediate clump destruction by the turbulence or to gravitational collapse before the action of the combined forces (gravity, magnetic, pressure, and turbulence). This also explains why we have considered an initial temperature around 10 K, which is actually more appropriate for the final dense cores that we want to generate than for  the initial more diffuse clumps of the CNM whose typical temperatures are of the order of 50 to 100 K.  Within an isothermal approximation (i.e., without a more realistic treatment of the radiative cooling to release the internal energy or heating excess which naturally leads to temperature decrease),  the starting with temperatures around 100 K would lead to clump destruction by the turbulence before the interplay of the other forces occur. Nevertheless, the adopted initial conditions for the diffuse clumps still correspond to a fiducial parametric space compatible with observations, as indicated in Figure \ref{fig:testes} where we have superposed our  simulated  clumps both in the beginning and after their evolution to the observed core diagram of magnetic field versus column density taken from Crutcher (2012).

The initial mass in stars in the cloud clump $M_{\star}$ was also varied between $20-60$ M$_{\odot}$. They produce a spherical gravitational potential, $\Psi_{\star}(r)$, given by:

\begin{equation}
  \Psi_{\star}(r\leq r_{max})= - \frac{G M_{\star}}{(r+r_{\star})}
\end{equation}

\noindent where $r$ is the distance to the center of the computational box, and $r_{max}\sim 1.45$pc is a cut-off in the gravitational force to ensure the symmetry while using periodical boundaries, and $r_{\star}$ is a smoothing radius. Its value gives the characteristic length of the clump. We fixed $r_{\star} = 0.325$pc in order to limit the values of the gravitational force and prevent the system to become Parker-Rayleigh-Taylor unstable.

In order to follow the evolution of the system we solve numerically the ideal MHD equations:

\begin{equation}
  \frac{\partial \rho}{\partial t}+ \nabla \cdot (\rho {\bf v}) = 0
\end{equation}

\begin{equation}
\begin{split}
   \rho\left(\frac{\partial}{\partial t} + {\bf v} \cdot \nabla\right){\bf v} =  & - c_s^2\nabla\rho + \frac{1}{4 \pi} (\nabla\times {\bf B} )\times {\bf B} \\
  & - \rho \nabla (\Psi_{gas} + \Psi_{\star}) + {\bf f}
\end{split}
\end{equation}

\begin{equation}
  \frac{\partial {\bf B}}{\partial t}= \nabla\times({\bf v}\times {\bf B})
\end{equation}

\noindent where the independent variables have their usual definition. The magnetic field ${\bf B}$, satisfies the divergence free condition $\nabla \cdot {\bf B} =0$.  The potential $\Psi_{gas}$ is due to the gas self-gravity which obeys the Poisson equation:

\begin{equation}
\nabla^2 \Psi_{gas}= 4 \pi G \rho
\end{equation}

The source term ${\bf f}$ of the second equation is the bulk force responsible for the injection of turbulence. {\bf We employ an isotropic, non-helical, solenoidal, delta correlated in time forcing $f$.} It induces harmonic velocity fields isotropically distributed in the Fourier space, concentrated around a typical wavelength that defines the injection scale $l_{inj}$. The power supply is constant, keeping the random velocity of the gas $v_{turb}$ approximately constant. In all our models with injection of turbulence, we use $l_{inj} \approx 1.3$ pc and $v_{turb}$ between $2.4 \times 10^{4}$ and $5.7 \times 10^{4}$ cm/s. The injection of turbulent energy starts at $t = 0.0 $ Myr and grows slowly until it reaches its maximum at $t \sim 10.6 $ Myrs. 

{\bf We employed a modified version of the} Cartesian Godunov-MHD {\bf code originally developed by G. Kowal} \citep{Kowal2007, Diego2008, Reinaldo2010}, using the HLL solver and Runge-Kutta of second order for time integration. The Poisson equation is solved by a multigrid based algorithm (\citet{nr1992}). The code is available upon request directly to the authors.


\subsection{Initial Conditions}

Table \ref{tab:non_equi} shows the initial conditions for all the simulated models both with and without self-gravity. Models without self-gravity are labelled  "R", the self-gravitating ones starting out of equilibrium are labelled "N"  and the one starting in magneto-hydrostatic equilibrium is labelled "E" model. The "R" models,  though scale independent, are also described  here by their initial conditions in physical unities for comparison with the other models, except for the non-dimensional parameter $A$ which measures the strength of the gravitational potential $A=(G M_{\star})/(L.c_s^2)$, where G is the Newton gravitational constant, $M_{\star}$ is the stellar mass potential, L is the size of the computational domain (which also gives  the distance unity in the code), and $c_s$ is the sound speed (which is also the velocity  unity in the code). The other parameters of the Table are the total mass $M_{tot}$, i.e., the sum of the mass in stars and the mass in gas, the initial cloud gas numerical density $n_0$, the initial magnetic field $B_0$, the thermal to magnetic pressure ratio $\beta$, the turbulent velocity $v_{\text{turb}}$, the Alfvénic velocity $v_{A}$, the initial cloud Jeans mass in the presence of magnetic field $m_{J,B}$, and the initial turbulent to magnetic energy density ratio $E_{turb}/E_{mag}$ for the core ($r_c\le 0.3$pc). The turbulent and Alfvén velocities are given in units of the isothermal sound speed $c_s = 3\times10^4$cm/s, which was chosen as the velocity unity in the code (see below).

\begin{table*}[!htb]\footnotesize
 \begin{center}
 \caption{Initial conditions for the models}
 \vspace{1.5ex}
 \centering
 \begin{tabular}{l c c c c c c c c c}
	\hline \hline\\
       [-1.0ex]
	  Model  & $M_{\star}$(M$_{\odot}$)  &    $A$    &  $n_0$(cm$^{-3}$)  &  $B_0$($\mu$G) & $\beta$ &  $v_{turb}$($c_s$)  &  $v_{A}$($c_s$)  &  $m_{J,B}$(M$_{\odot})$   &   $E_{turb}/E_{mag}$ \\
      [0.5ex]	\hline \\   [-1.2ex]
	  R1     &   $61.1$    &      $0.9$    &     $10$     &   $0.35$    &  $3.0$  &  $0.8$  &  $0.8$  &    $373.7$   &  2.96  \\
	  R2     &   $40.7$    &      $0.6$    &     $90$     &   $1.06$    &  $3.0$  &  $0.8$  &  $0.8$  &    $314.1$   &  2.90  \\
	  R3     &   $27.1$    &      $0.4$    &    $100$     &   $1.12$    &  $3.0$  &  $0.8$  &  $0.8$  &    $326.6$   &  2.89  \\
      [1ex] \hline \\   [0.5ex]
	  Model  &  $M_{\star}$(M$_{\odot}$)  &  $M_{tot}$(M$_{\odot})$  &  $n_0$(cm$^{-3}$)  &  $B_0$($\mu$G) & $\beta$ &  $v_{turb}$($c_s$)  &  $v_{A}$($c_s$)  &  $m_{J,B}$(M$_{\odot})$   &   $E_{turb}/E_{mag}$ \\
      [0.5ex]
	\hline \\
      [-1.2ex]
	  N1     &   $61.1$    &     $69.5$    &     $10$     &   $0.35$    &  3.0  &  0.8  &  0.8  &    $373.7$   &  2.96  \\
	  N2a    &   $40.7$    &     $49.1$    &     $10$     &   $0.35$    &  3.0  &  0.8  &  0.8  &    $448.1$   &  2.96  \\
	  N2b    &   $40.7$    &    $116.4$    &     $90$     &   $1.06$    &  3.0  &  0.8  &  0.8  &    $314.1$   &  2.90  \\
	  N2c    &   $40.7$    &    $116.4$    &     $90$     &   $1.85$    &  1.0  &  1.2  &  1.4  &    $379.8$   &  2.15  \\
	  N2d    &   $40.7$    &    $116.4$    &     $90$     &   $3.37$    &  0.3  &  1.9  &  2.6  &    $639.7$   &  1.62  \\
	  N2e    &   $40.7$    &    $108.0$    &     $80$     &   $1.74$    &  1.0  &  1.2  &  1.4  &    $389.9$   &  2.16  \\
	  N3     &   $27.1$    &    $111.2$    &    $100$     &   $1.12$    &  3.0  &  0.8  &  0.8  &    $326.6$   &  2.89  \\
	  N4     &   $20.4$    &    $104.5$    &    $100$     &   $1.12$    &  3.0  &  0.8  &  0.8  &    $339.6$   &  2.89  \\
	  E1     &   $40.7$    &     $46.6$    &     $90$     &   $1.06$    &  3.0  &  0.8  &  0.8  &    $390.1$   &  2.90  \\
	\hline
 \end{tabular}
 \label{tab:non_equi}
 \end{center}
\end{table*}

Table \ref{tab:times} shows the dynamical time, $t_{dyn}=L/c_s$. This is compared with the free-fall time of the cloud which is given by,

\begin{equation}
  t_{ff} = \sqrt{\frac{3\pi}{32\;G\;\rho}}
\end{equation}

\noindent where $\rho$ is the total density corresponding to the total mass, $M_{tot}$(M$_{\odot})$ as in Table \ref{tab:times}, and the reconnection diffusion time of the magnetic field, $t_{\rm diff}$, 

\begin{equation} 
 t_{diff}=\frac{l^2}{\eta}, 
 \label{eq:diff_time}
\end{equation}

\noindent where the reconnection diffusivity of the gas is (Santos-Lima et al. 2010; Lazarian 2006; 2011; Lazarian et al. 2012b):

\begin{equation}
\begin{split}
&\eta \sim l_{\rm inj}v_{\rm turb} \ \  \  \  \  \  \ \ \ \ \ \ \ \ \;  {\rm if}\;  v_{\rm turb} \geq v_A\; ,\\ 
&\eta\sim l_{\rm inj}v_{\rm turb} \left(\frac{v_{\rm turb}}{v_A}\right)^3\; \; {\rm if}\; v_{\rm turb} < v_A\; ,
\end{split} 
\label{eq:diffusivity}
\end{equation}

\noindent where $l_{inj}=L/k_f$. The relations above indicate that the ratio $\left(v_{\rm turb}/v_A\right)^3\;$ is important only in a regime of sub-Alfv\'enic turbulence, i.e. with the  Alfv\'enic Mach number  $M_A \leq 1$. In \ref{eq:diff_time} above, $l$ is a characteristic scale smaller than L where  the magnetic field is diffused. In order to compare the models, in all turbulent simulations we considered the same  $\eta$. Thus for a given characteristic scale $l$, when changing $\beta$ we also changed $v_{turb}$ in order to keep $\eta$ constant. This  ensured  that the turbulence decayed  at approximately the same time scale in all models. In table \ref{tab:times}, the diffusion time scale was computed considering  the cloud core scale, i.e., $l \simeq 0.3$  pc. 

\begin{table*}[!htb]
 \begin{center}
 \caption{Characteristic timescales.}
 \vspace{1.5ex}
 \centering
 \begin{tabular}{l c c c c}
	\hline	\hline \\
      [-1.0ex]
	  Model & $t_{dyn}$(Myrs) & $t_{ff}$(Myrs)  &  $t_{diff}$(Myrs) \\
	[0.5ex]
       \hline \\
      [-1.0ex]
	  R1    &    $10.6$     &   $5.7$       &    $1.4$       \\     
	  R2    &    $10.6$     &   $4.4$       &    $1.4$       \\     
	  R3    &    $10.6$     &   $4.5$       &    $1.4$       \\ 
	[0.5ex]
       \hline \\
      [-1.0ex]
	  N1    &    $10.6$	   &    $5.7$	&    $1.4$	 \\
	  N2a   &    $10.6$    &    $6.7$   	&    $1.4$   \\
	  N2b   &    $10.6$    &    $4.4$   	&    $1.4$   \\
	  N2c   &    $10.6$    &    $4.4$   	&    $1.4$   \\
	  N2d   &    $10.6$    &    $4.4$   	&    $1.4$   \\
	  N2e   &    $10.6$    &    $4.5$   	&    $1.4$   \\
	  N3    &    $10.6$    &    $4.5$   	&    $1.4$	 \\
	  N4    &    $10.6$    &    $4.6$   	&    $1.4$	 \\
	  E1    &    $10.6$    &    $5.6$   	&    $1.4$	 \\
	 \hline
 \end{tabular}
 \label{tab:times}
 \end{center}
\end{table*}

Also, in order to be able to detect the transport of magnetic flux during the dynamical evolution of the collapsing clouds, we made the initial reconnection diffusion time smaller than the initial free-fall time for all simulated models. We followed the evolution of $B$ and $\rho$ for at least ten time steps, corresponding to ten dynamical times, or in total $\sim 100$ Myrs. For comparison with the turbulent MHD {\bf models, we also performed simulations with similar initial conditions,} but without injecting forced turbulence (these will be hereafter referred as "laminar" or non-turbulent models).


\section{Results of the numerical simulations}
\label{Results}

\subsection{Models with self-gravity}

We present here the results for nine models where we have included the effects of self-gravity for which the initial conditions are listed in Table \ref{tab:non_equi}. All models, but N2c, N2d, and N2e, have a turbulent velocity which is equal to the initial Alfvén speed, while these three models are sub-Alfvénic, i.e., have a turbulent velocity smaller than the Alfvén speed. The non-equilibrium models N1 and N2a were initialized with the same gas density $ \rho = 1.67 \times 10^{-23}$g/cm $^{3}$, but different gravitational potentials (i.e., different values of the stellar mass potential $M_{\star}$). Model N2b has the same initial stellar potential as in model N2a, $40.7$M$_{\odot}$, but a larger gas density $ \rho = 1.503 \times 10^{-22}$g/cm $^{3}$ which allows to compare the effects of self-gravity. Models N2c and N2d have the same initial gas density as in the previous model, N2b, but different values of $\beta$  which allow to compare the effects of the magnetic field. Model N2e has the same initial $\beta$ as in model N2d but a slightly smaller gas density $ \rho = 1.336 \times 10^{-22}$g/cm $^{3}$  which allows turbulence to be more dominant than in the previous case. The models N3 and N4 have the same initial  $\beta = 3.0$ and gas density $ \rho = 1.67 \times 10^{-22}$g/cm $^{3}$, but different stellar gravitational potentials, $27.1$M$_{\odot}$ and $20.4$M$_{\odot}$, respectively. Both have smaller stellar gravitational potentials but larger gas density (therefore more significant self-gravity) than the previous models. Finally, the model starting in equilibrium E1 has conditions which are similar to those of  model N2b. The remaining initial conditions are the same for all models.

\begin{figure*} [!htb]
		\includegraphics[width=0.98\linewidth]{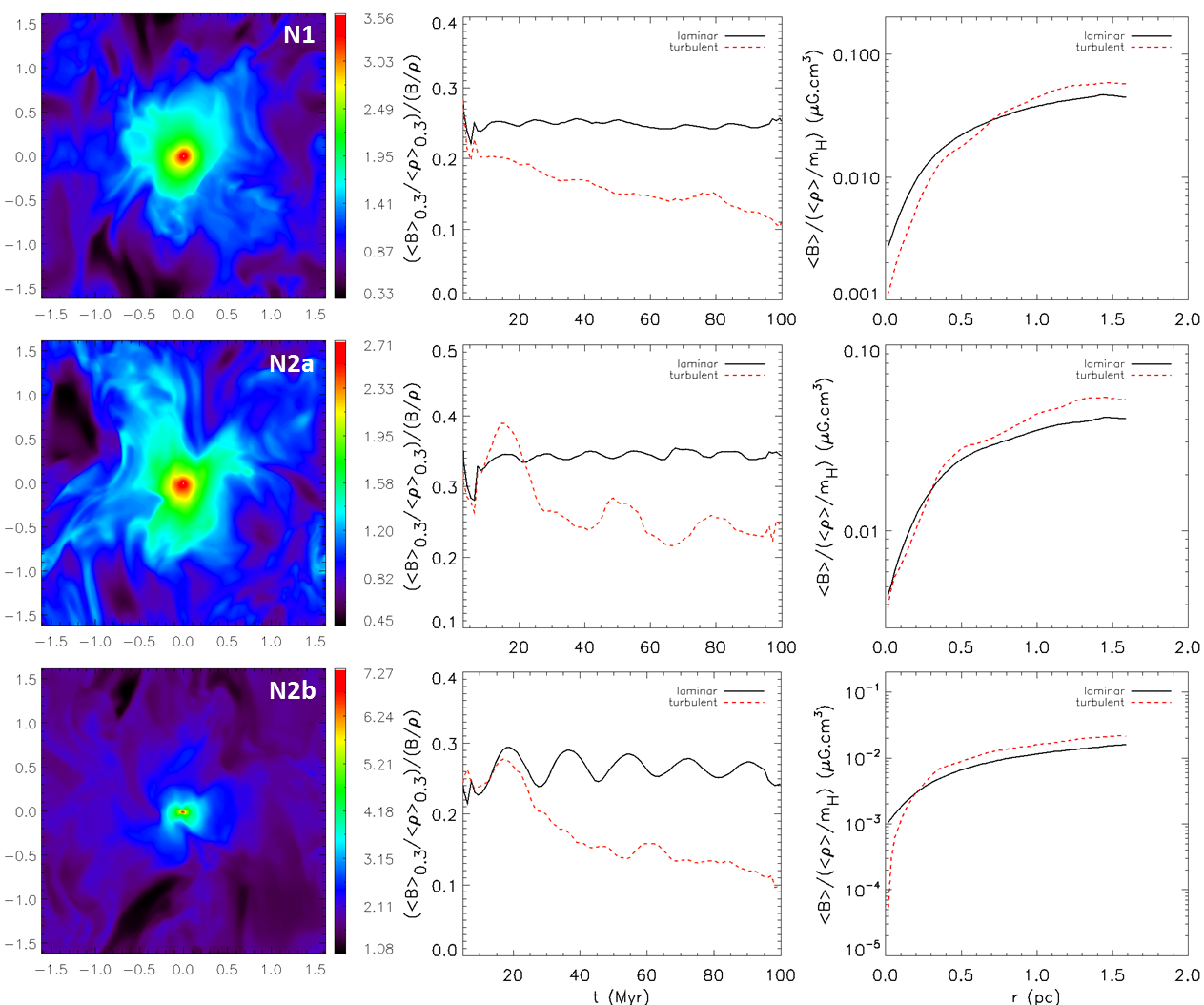}
  \caption{Model N1 with $M_{pot}=61.1$M$_{\odot}$ (top), models N2a (center) and N2b (bottom) with $M_{pot}=40.7$M$_{\odot}$. Top and middle models have initial densities  $ n_0 =10 $ cm$^{-3}$, and the bottom model N2b has $n_0 = 90$ cm$^{-3}$. All models have $\beta = 3.0$. Left panels show logarithmic density maps of the central slices of the turbulent cloud models at $t = 100$ Myrs. The axis of these maps are in parsecs with the origin in the center of the computational domain. Middle panels show the temporal evolution of the average magnetic field-to-density ratio at the cloud core region of radius $r_c= 0.3$ pc normalized by the average value over the entire cloud, $(\langle B \rangle_{0.3} / \langle \rho \rangle_{0.3})/(\bar{B}/\bar{\rho})$, for the turbulent (red-dashed lines) and the laminar (black continuous lines) models. Right panels show the radial profile of the average magnetic field-to-density ratio at $t = 100$ Myr $ \langle B\rangle / \langle \rho \rangle $ for these models.  \textbf{ In the turbulent models, the injection of turbulent energy starts at $t = 0.0$ Myr and grows slowly until it reaches a maximum around $t \sim 10$ Myr. Although most of the models in this work have been evolved for $\sim$100 Myr, we find that an effective diffusion of magnetic flux in the turbulent models takes place  around 30 Myr already, i.e., approximately 20 Myr after the full development of the turbulence (see more details in the text). Adapted from de Gouveia Dal Pino et al. (2012).}}
\label{fig:g0.6}
\end{figure*}

The left panels of Figures \ref{fig:g0.6}, \ref{fig:g0.6b}, and \ref{fig:g0.4}, and top-left panel of Figure \ref{fig:equil}, {\bf present maps of the logarithmic density of the central slices of the cloud models after $\sim$ten dynamical times ($\sim 100$ Myr).}

\textbf{In the middle panels of these figures, we compare the time evolution of the average magnetic field-to-density ratio inside the core of the cloud  with radius $r=0.3$ pc,  $(\langle B\rangle_{0.3}/\langle \rho\rangle_{0.3})/(\bar{B}/\bar{\rho})$, for the turbulent (red-dashed lines) and the laminar (black continuous lines) models.} The brackets mean averages taken inside the cloud radius $0.3$ pc. In this averaging, only the z component of the magnetic field has been considered. This ratio is normalized by the average value $(\bar{B}/\bar{\rho})$ taken over all the computational box, where $\bar{B}$ also stands for the z component of the magnetic field only. For the adopted geometry with an initial uniform magnetic field along the z direction, the ratio above is approximately equal to the magnetic flux-to-mass ratio within the region considered.    

The right panels of these figures (\ref{fig:g0.6}, \ref{fig:g0.6b}, and \ref{fig:g0.4}) and the top-right panel of Figure \ref{fig:equil} depict profiles of the average magnetic field-to-density ratio as a function of the radial distance $r$ from the center of the cloud for the same models, at $t= 100$ Myr. 

\begin{figure*} [!hbt]
		\includegraphics[width=0.98\textwidth]{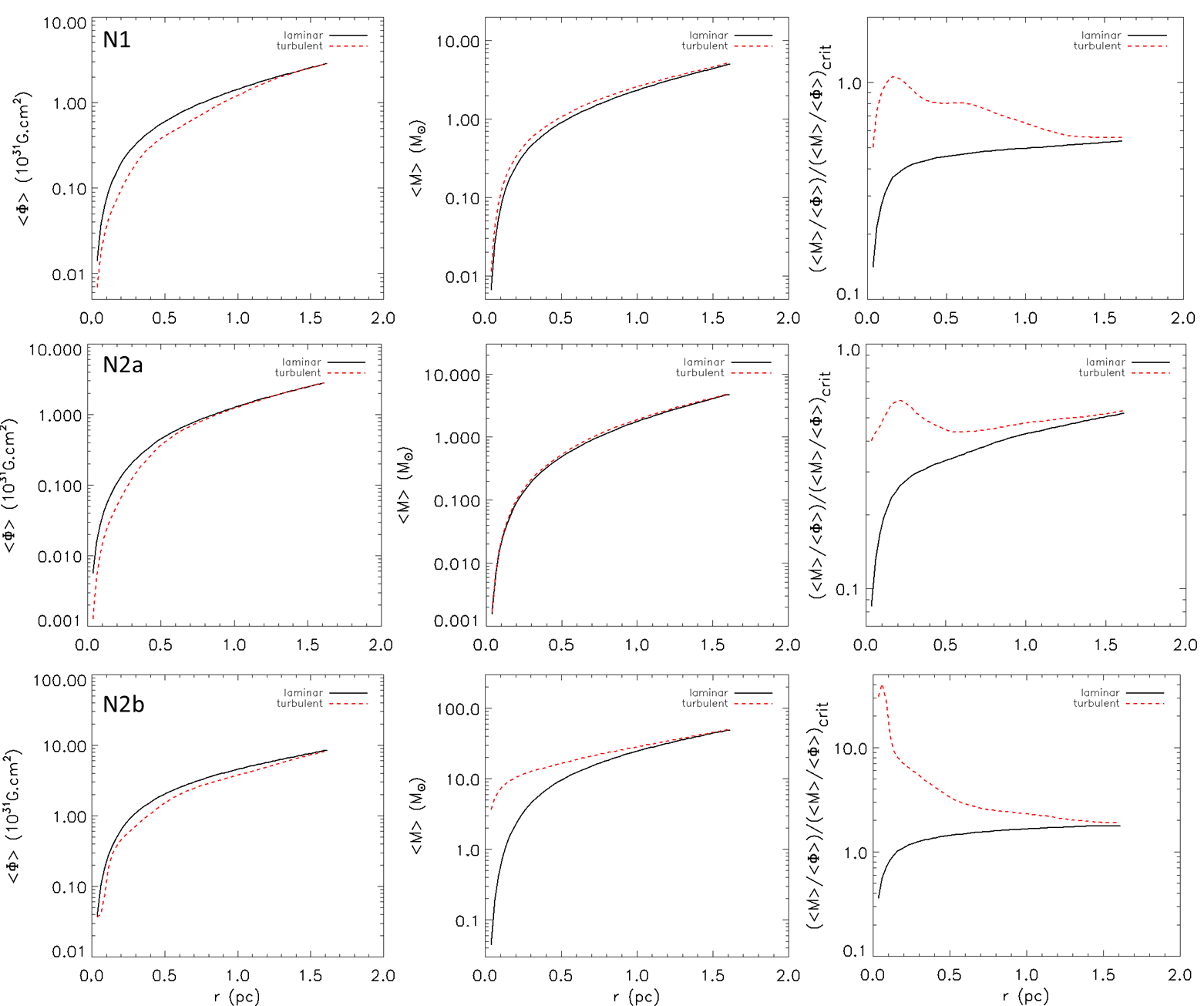}
  	\caption{Radial profiles at $t= 100$ Myr for the magnetic flux $\Phi$ (left panels), the gas mass $M$ (center panels), and  for the mass-to-magnetic flux ratio normalized by the critical value $(M/\Phi)/(M/\Phi)_{crit}$ (right panels) for the models of Figure \ref{fig:g0.6}. Top panels: model N1 ($n_0 = 10$ cm$^{-3}$, and $M_{\star}=61.1$M$_{\odot}$); center panels: model N2a ($n_0 = 90$ cm$^{-3}$, and $M_{pot}=40.7$M$_{\odot}$); bottom panels: model N2b ($ n_0 = 10$ cm$^{-3}$). Red-dashed lines are for turbulent models and black continuous lines are for the laminar models.}
\label{fig:MtoFInteg}
\end{figure*}

In Figure \ref{fig:g0.6}, \textbf{the laminar models show a nearly constant  magnetic field-to-density ratio inside the cloud cores   after a fast decrease at the beginning caused by the relaxation of the system (middle panels). The turbulent models, on the other hand, clearly show a decrease of this ratio}\footnote{{\bf We note that the oscillations observed in these plots (which are slightly stronger in the laminar models) are acoustic oscillations of the cloud} and occur because {\bf the virialization time of these systems is longer than the simulated period} (see also de Gouveia Dal Pino et al. 2012).}. {\bf This result suggests that magnetic flux has been transported from the denser internal regions to the outer, less dense regions of the cloud clumps.} \textbf{ This effect is stronger in model N2b (bottom panel) which has initial gas density much larger than the other models and therefore, is under the influence of larger self-gravity.} \textbf{The comparison of models N1 and N2a (which have the same initial gas density) indicates that the one with larger stellar potential (model N1) exhibits a larger decoupling between the magnetic flux and the mass density.} Consistently, the right panels of Figure \ref{fig:g0.6} show that the turbulent model with larger initial density (for which the effect of  self-gravity is more important, i.e., model N2b), has a larger decrease in the radial profile of the magnetic field-to-density ratio in the core region (accompanied by a larger increase in the outer regions of the cloud) than the other turbulent models, therefore confirming the trend detected in the middle panels (see also de Gouveia Dal Pino et al. 2012).

In order to verify if the initially subcritical clouds have developed supercritical cores after the action of the reconnection transport, Figures \ref{fig:without02}, \ref{fig:MtoFInteg}, \ref{fig:MtoF3Integ} and \ref{fig:MtoF2Integ}, and the bottom panels of Figure \ref{fig:equil} compare the radial profiles of the integrated magnetic flux (left panels), gas mass (center panels), and mass-to-magnetic flux ratio normalized by the critical value (right), at $t=100$ Myr for the models with and without turbulence\footnote{The critical value of the mass-to-magnetic flux ratio for a disk shaped system is given by: $(2\pi \rm{G}^{1/2})^{-1}$, Crutcher (2005b)}. For instance, Figure \ref{fig:MtoFInteg} indicates that the models N1 and N2b of Figure \ref{fig:g0.6} develop supercritical cores, while N2a remains subcritical.

\begin{figure*} [!htb]
\centering
	\includegraphics[width=0.98\linewidth]{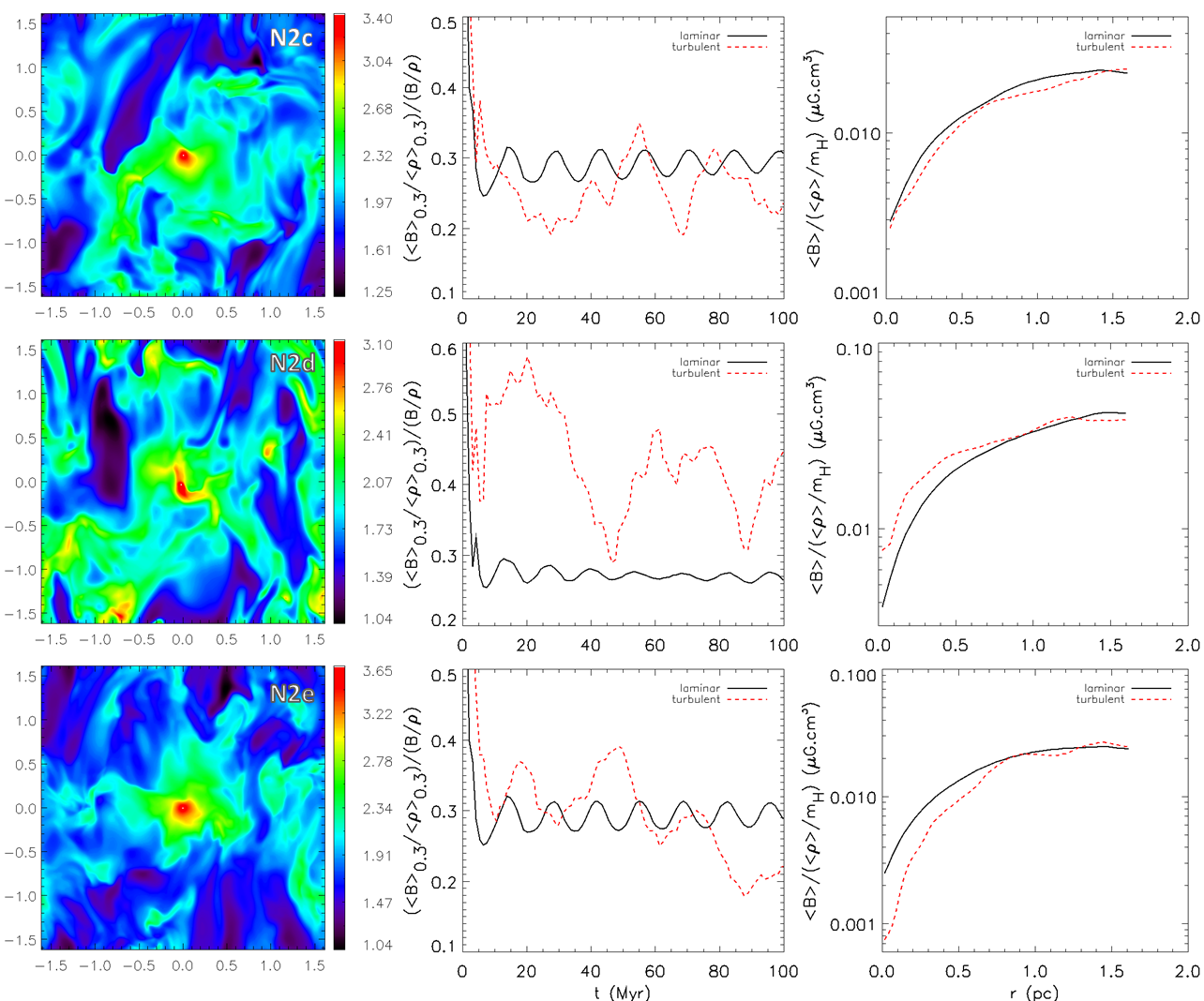}
  \caption{The same as in Figure \ref{fig:g0.6} for models N2c, N2d, and N2e. Top and Bottom: models N2c and N2e with $\beta=1.0$ Center: model N2d  with $\beta=0.3$. Top and middle models have initial densities $n_0 = 90$ cm$^{-3}$, and the bottom model has $n_0 = 80$ cm$^{-3}$. All models have $M_{pot}=40.7$M$_{\odot}$}
\label{fig:g0.6b}
\end{figure*}

\begin{figure*} [!htb]
\centering
	\includegraphics[width=0.98\textwidth]{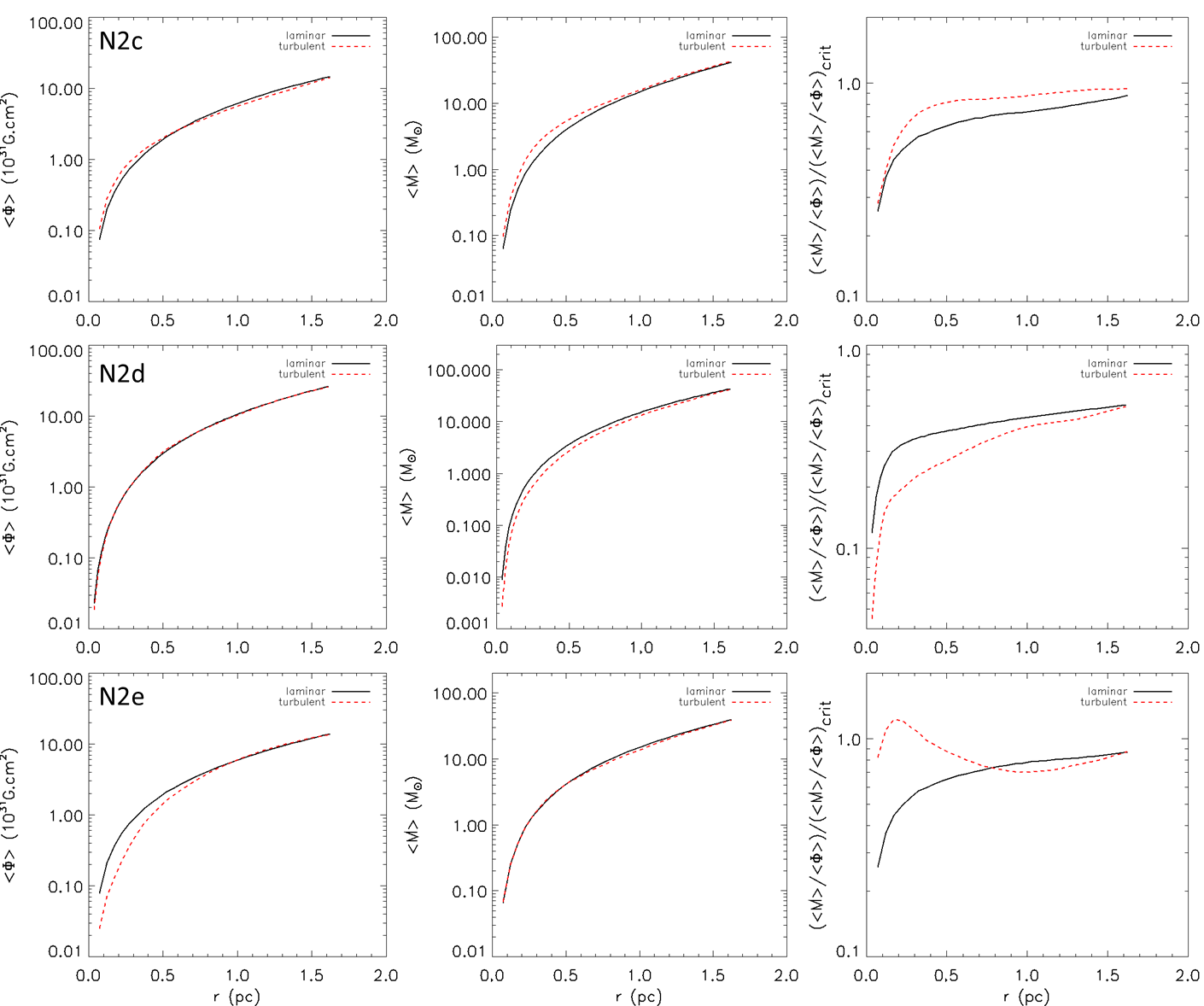}
  \caption{The same as in Figure \ref{fig:MtoFInteg} for models N2c, N2d, and N2e. Top panels: model N2c ($n_0 = 90$ cm$^{-3}$, and $\beta=1.0$); center panels: model N2d ($n_0 = 90$ cm$^{-3}$, and $\beta=0.3$); bottom panels: model N2e ($ n_0 = 80$ cm$^{-3}$, and $\beta=1.0$).}
\label{fig:MtoF3Integ}
\end{figure*}

Figures \ref{fig:g0.6b} and \ref{fig:MtoF3Integ} show the density maps and similar profiles as those of Figure \ref{fig:g0.6} for the models with sub-Alfvénic turbulence N2c, N2d, and N2e. Models N2c and N2d have the same initial gas density and stellar mass as in model N2b, but smaller initial $\beta$ and turbulent to Alfvén velocity ratio. This naturally makes the decoupling of the magnetic flux from the dense gas regions by reconnection diffusion more difficult. Both models N2c and N2d, for which $\beta=1.0$, and $0.3$, respectively, have no significant magnetic flux transport outward when compared to model N2b in Figure \ref{fig:g0.6} for which $\beta =3.0$, or to their laminar counterparts. However, when the initial mass density of model N2c is decreased from 90 to 80 cm$^{-3}$ (keeping $\beta= 1.0$), as in model N2e, the transport of magnetic flux by turbulence is slightly  enhanced (though still much less than in model N2b) and the initially subcritical cloud develops a nearly critical core after 100 Myr (see Figure \ref{fig:MtoF3Integ}). This result was also confirmed in higher resolution ($512^3$) simulations of model N2e.

Therefore, this result indicates that the increase of gas density and total gravity will not always result in enhancement of the flux transport by the turbulence as we have seen in the models of Figure \ref{fig:g0.6}. When turbulence is sub-Alfvénic (and thus the magnetic field is strong for the level of turbulence applied), a smaller gravitational potential will delay the collapse and thus may give time for the reconnection diffusion to transport outward part of  the magnetic flux. This explains why the decrease in density (and this of self-gravity) from $n_0=90$ cm$^{-3}$ in model N2c to $n_0=80$ cm$^{-3}$ in model N2e, results in the build up of a nearly critical core in the last case (see also discussion in Sections 4.1 and 4.2).      

\begin{figure*} [!htb]
\centering
	\includegraphics[width=0.98\linewidth]{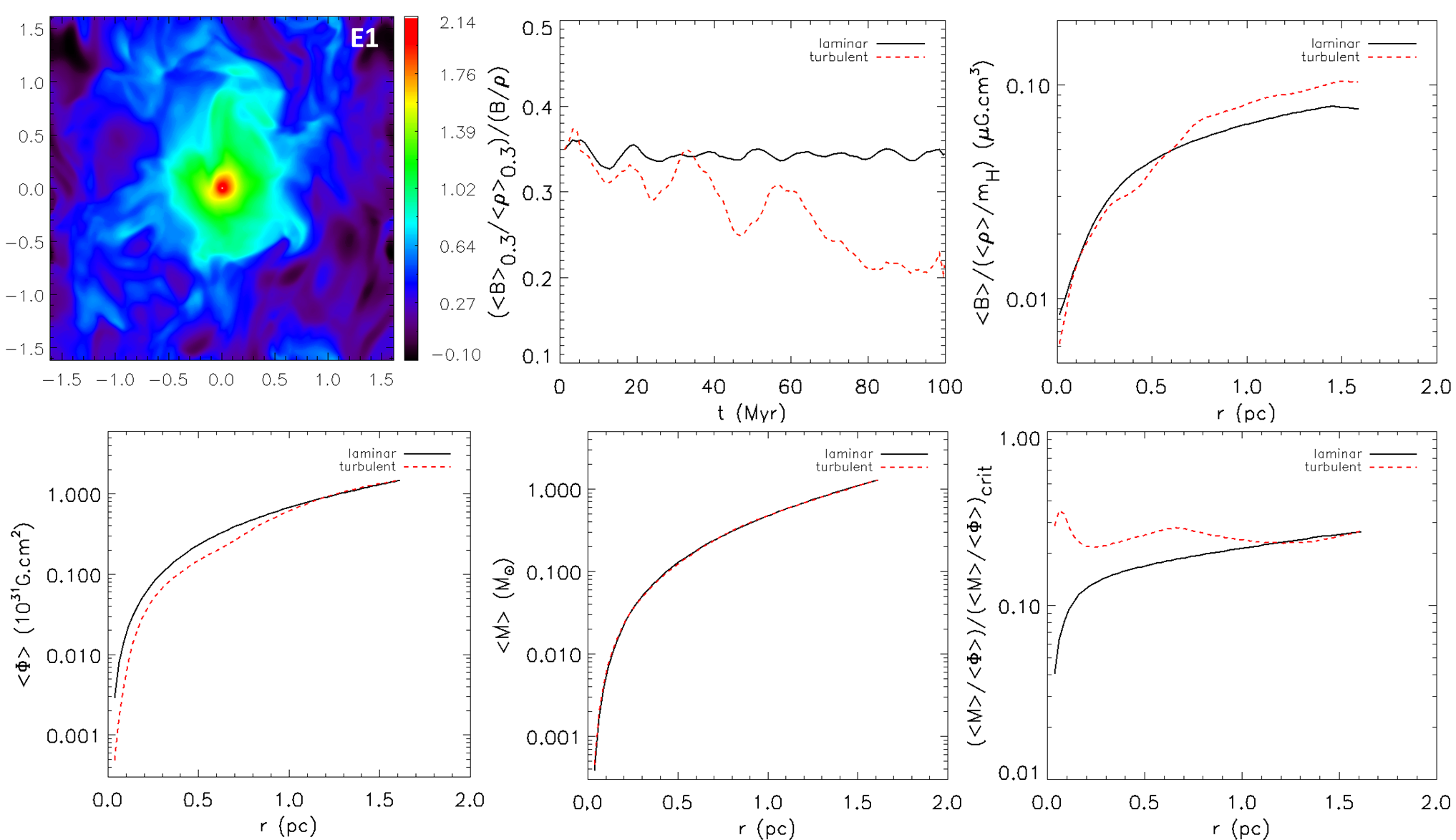}
  \caption{Equilibrium model E1 with $M_{pot}=61.1$M$_{\odot}$. This model has initial central density $n_0 = 90 $ cm$^{-3}$ and $\beta = 3.0$. Top: the same as in Figure \ref{fig:g0.6} for model E1. Bottom: the same as in Figure \ref{fig:MtoFInteg} for model E1.}
\label{fig:equil}
\end{figure*}

Figure \ref{fig:equil} shows the density map and profiles for model E1. This model which starts in magneto-hydrostatic equilibrium has the same initial $\beta$  and density $\rho_0$ in its central region as the non-equilibrium model N2b, but the equilibrium condition makes its density profile stratified and thus the total mass inside the cloud is smaller than in N2b model (see Table \ref{tab:non_equi}). As in the turbulent N2b model, E1 also suffers outward magnetic flux transport, as indicated by the decreasing magnetic field-to-density ratio in the core region (top-middle panel). However, the much smaller total mass of E1 prevents it to collapse and become a supercritical core, as indicated by the mass-to-magnetic flux diagrams of Figure \ref{fig:equil} (bottom-right panels, red-dashed line).

\begin{figure*} [!htb]
\centering
	\includegraphics[width=0.98\textwidth]{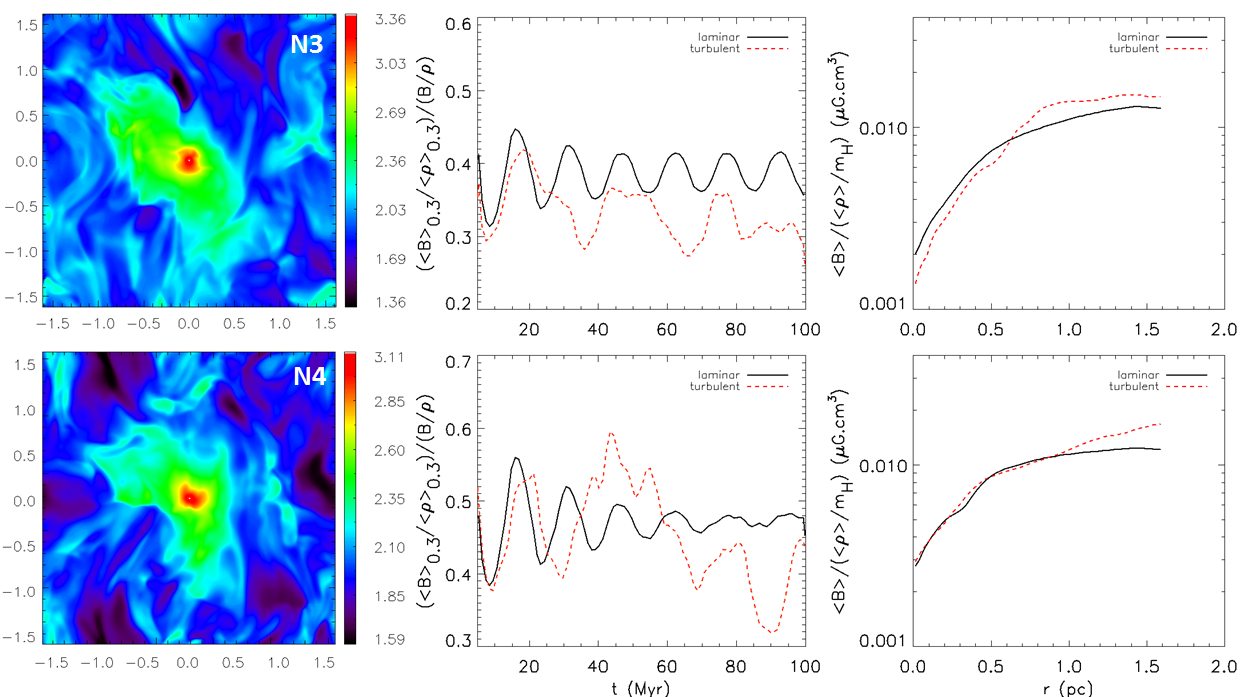}
  \caption{The same as in Figure \ref{fig:g0.6} for the non-equilibrium models N3 and N4. Top: model N3 with $M_{pot}=27.1$M$_{\odot}$ bottom: model N4 with $M_{pot}=20.4$M$_{\odot}$. Both models have initial densities $n_0 = 100$ cm$^{-3}$ and $\beta = 3.0$.}
\label{fig:g0.4}
\end{figure*}

\begin{figure*} [!htb]
\centering
 \includegraphics[width=0.98\textwidth]{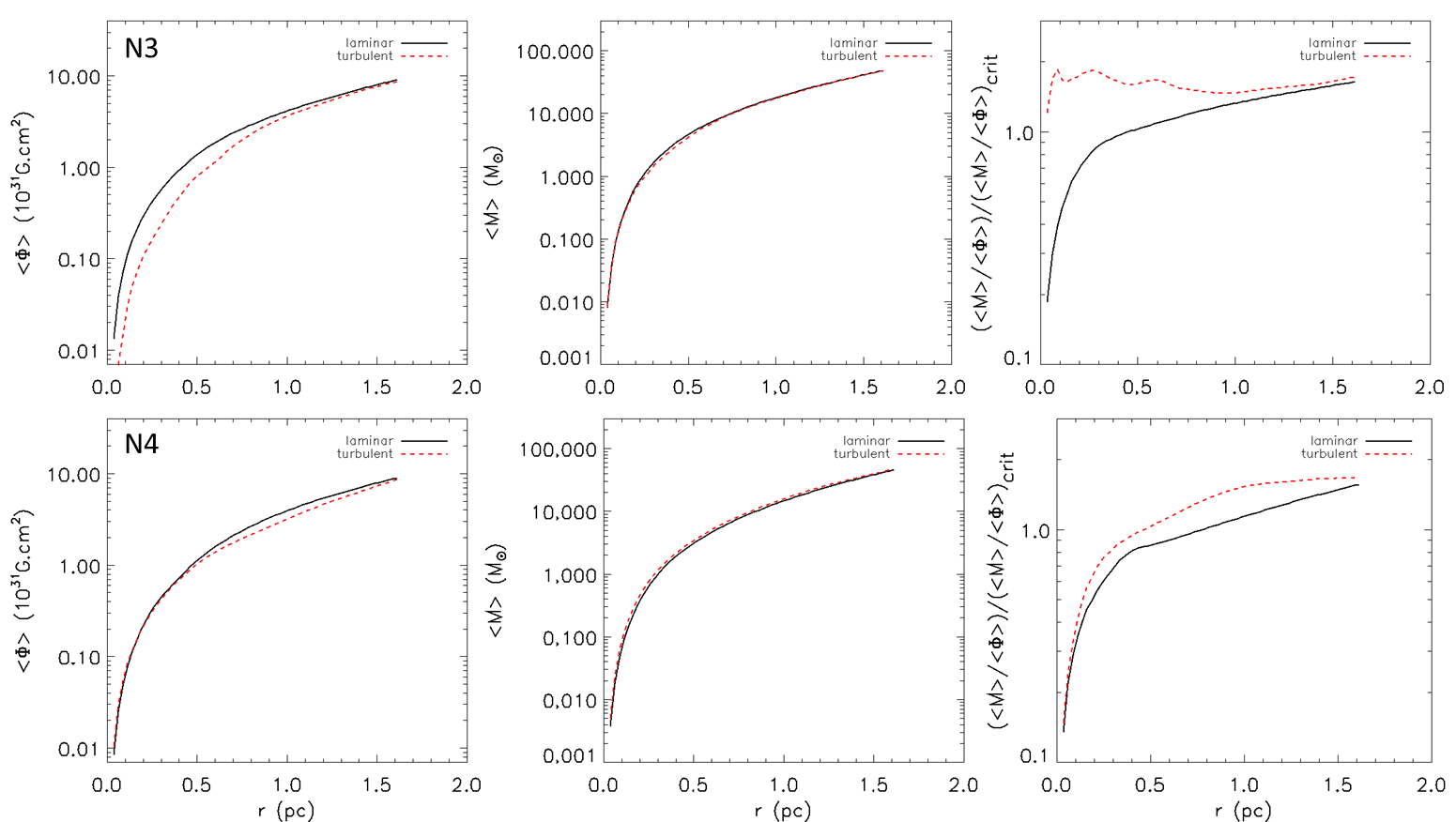}
 \caption{The same as in Figure \ref{fig:MtoFInteg} for models N3 and N4 of Figure \ref{fig:g0.4}. Top panels: model N3 ($n_0 = 100$ cm$^{-3}$, and $M_{\star}=27.1$M$_{\odot}$); bottom panels: model N4 ($n_0 = 100$ cm$^{-3}$, and $M_{\star}=20.4$M$_{\odot}$).}
 \label{fig:MtoF2Integ}
\end{figure*}

Figures \ref{fig:g0.4} and \ref{fig:MtoF2Integ} show the results for models N3 and N4 which have the same initial $\beta = 3.0$ and gas density $ n_0= 100$ cm$^{-3}$ (which are comparable with model N2b for which the initial gas density is slightly smaller), but different stellar gravitational potentials, $27.1$M$_{\odot}$ and $20.4$M$_{\odot}$, respectively (which are smaller than that of model N2b).

In both turbulent models we note a smaller decrease with time of the magnetic field-to-density ratio in the cloud core relative to model N2b (see middle panels of Figure \ref{fig:g0.4}), specially for the model with smaller stellar potential (N4). The inspection of the radial profile of the mass-to-magnetic flux ratio at $t= 100$ Myr for this model in Figure \ref{fig:MtoF2Integ} shows that in fact, there is no transport of magnetic flux in this case. This occurs because of the combination of two effects. The larger magnetic field of model N4 relative to model N2b (see Table \ref{tab:non_equi}) makes it more difficult for the turbulence to decouple the magnetic flux from the denser material. At the same time, the smaller stellar gravitational well slows down the infall of matter to the center in model N4 (see Table \ref{tab:times}), while the action of the turbulence  helps to spread the gas against gravity (bottom left panel of Figure \ref{fig:MtoF2Integ}). Model N3 on the other hand, although with the same strength of magnetic field as model N4, it has a stellar potential large enough to push the core matter inwards while turbulence  decouples the magnetic flux transporting it to the less dense regions, allowing the formation of a supercritical core (as we see in the upper panels of Figure \ref{fig:MtoF2Integ}).


\subsection{Effects of Resolution on the Results}
\label{resolution}

\begin{figure}[!htb]
	\centering
		\includegraphics[width=7.5cm]{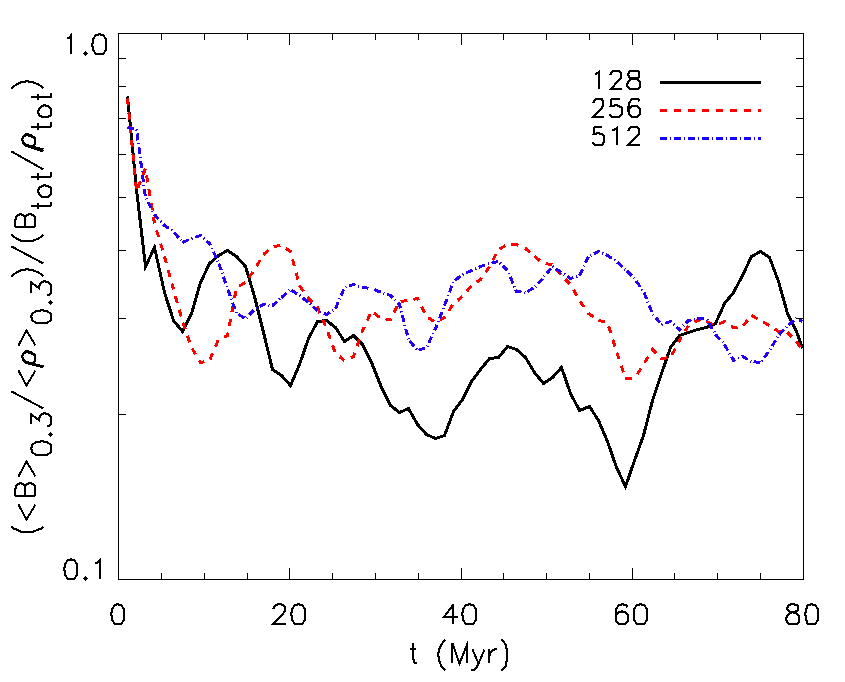}
	\caption{Comparison between different resolutions for the model N2e (Table \ref{tab:non_equi}).}
	\label{fig:resol}
\end{figure}

As remarked, all the numerical simulations presented above have been performed with a resolution $256^3$. 
In order to ensure {\bf that the results above are  not  affected by numerical effects, we  have also run  one of the models} (model N2e, Table \ref{tab:non_equi}) {\bf with $128^3$ and $512^3$ resolutions.} Fig. \ref{fig:resol} compares \textbf{the  evolution of the average magnetic field-to-density ratio at the core region (of radius 0.3 pc)   normalized by the average value over the entire cloud (as in Figure \ref{fig:g0.6b}, middle panel) for these three resolutions. }
 {\bf Since we do not observe significant differences between them, we can expect that the results presented for the models with resolution of $256^3$ are robust (see also the discussion section below).}

\subsection{Models with no self-gravity}

We have not discussed yet the specific effects that the inclusion of self-gravity produces  on the turbulent core collapse. In order to do that, we have to compare the self-gravitating models analysed in the previous section with counterparts $without$ self-gravity. These models labelled "R" are also listed in Table \ref{tab:non_equi}. Model R1 has similar initial conditions to those of the self-gravitating model N1 of Table \ref{tab:non_equi}, with a gravitational  potential parameter $A=0.9$ which is equivalent to a stellar mass of $61.7$ M$_{\odot}$. Models R2 and R3, in turn, have  similar initial conditions to the self-gravitating models N2b and N3 of Table 1, with  $M_{\star}=40.7$ M$_{\odot}\rightarrow A=0.6$, and $M_{\star}=27.1$ M$_{\odot}\rightarrow A=0.4$, respectively.

\begin{figure*} [!htb]
		\includegraphics[width=0.98\linewidth]{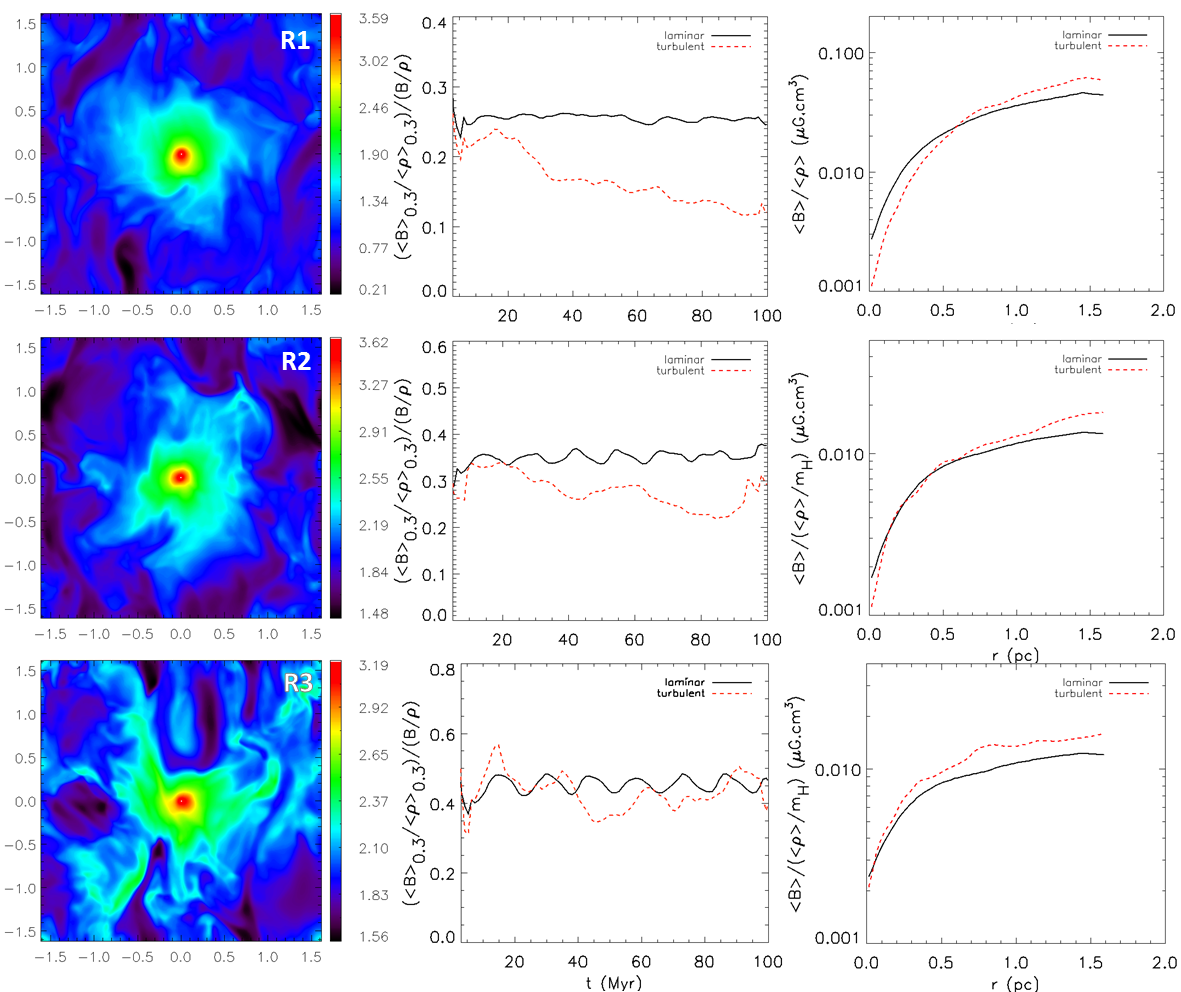}
  \caption{Model R1 with $A=0.9$ ($M_{pot}=61.1$M$_{\odot}$) (top), model R2 (center) with $A=0.6$  ($M_{pot}=40.7$M$_{\odot}$), and model R3 (bottom) with $A=0.4$ ($M_{pot}=27.1$M$_{\odot}$). The models have initial densities  $ \rho_0 =1.0 $, and  $\beta = 3.0$. Left panels show logarithmic density maps in the xz-plane of the {\bf central slices of the turbulent cloud models at $t = 100$ Myrs.  Middle panels show the temporal evolution of the average magnetic field-to-density ratio at the cloud core region of radius $r_c= 0.3$ pc normalized by the average value over the entire cloud, $(\langle B \rangle_{0.3} / \langle \rho \rangle_{0.3})/(\bar{B}/\bar{\rho})$, for the turbulent (red-dashed lines) and the laminar (black continuous lines) models.} Right panels show the radial profile of the average magnetic-to-density ratio at $t = 100$ Myrs $ \langle B\rangle / \langle \rho \rangle $ for these models.}
\label{fig:without01}
\end{figure*}

\begin{figure*} [!hbt]
		\includegraphics[width=0.98\textwidth]{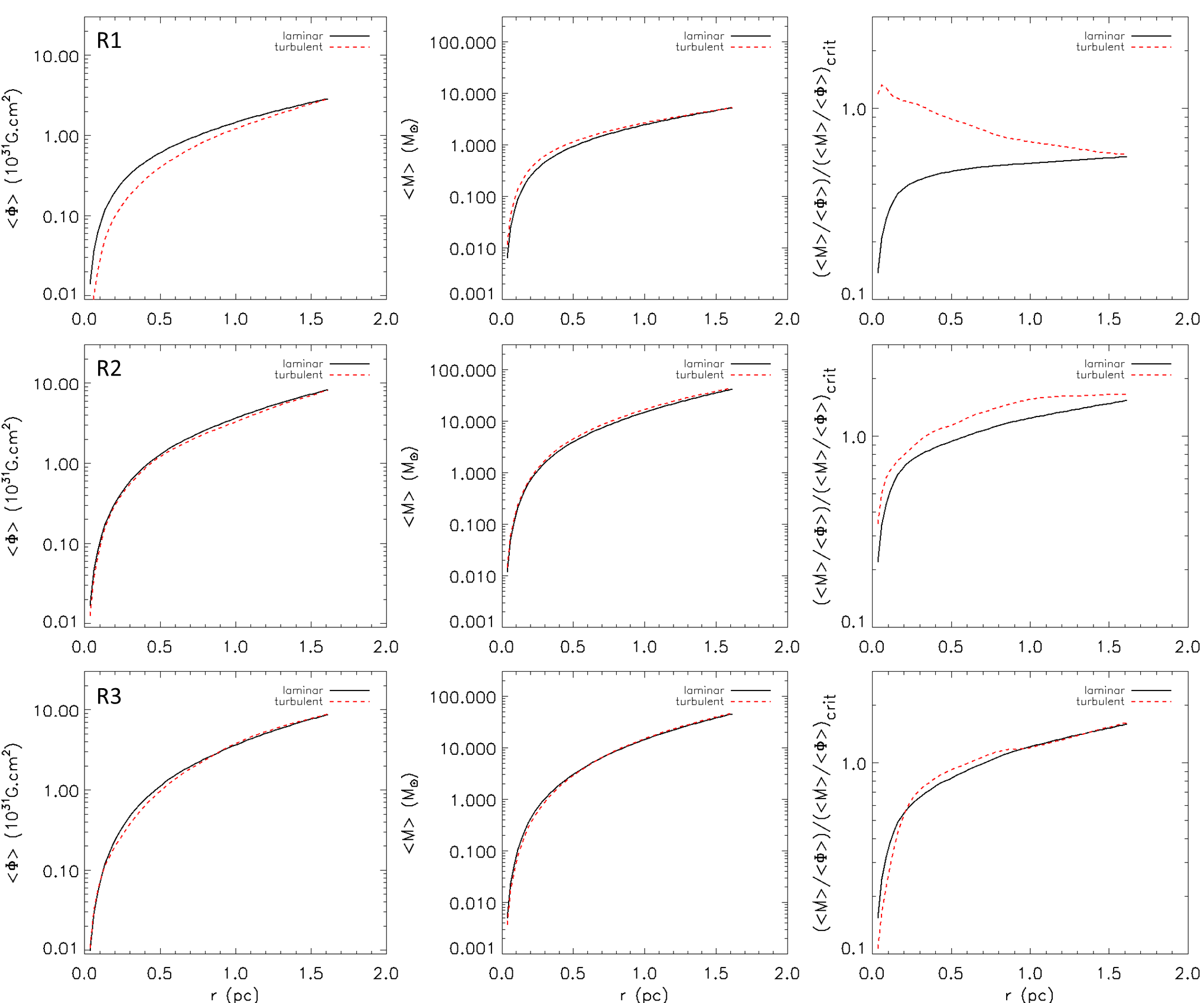}
  	\caption{Radial profiles at $t= 100$ Myr for the magnetic flux $\Phi$ (left panels), the gas mass $M$ (center panels), and the magnetic flux-to-mass ratio  normalized by the critical value $(M/\Phi)/(M/\Phi)_{crit}$ (right panels) for the models of Figure \ref{fig:without02}. Top panels: model R1 ($A=0.9$, i.e. $M_{\star}=61.1$M$_{\odot}$); center panels: model R2 ($A = 0.6$ , i.e. $M_{\star}=41.7$M$_{\odot}$); bottom panels: model R3 ($A = 0.4$ , i.e. $M_{\star}=27.1$M$_{\odot}$). Red-dashed lines are for turbulent models and black continuous lines are for the laminar models.}
\label{fig:without02}
\end{figure*}

The logarithmic density maps at 100 Myrs as well as all the related radial and time evolution profiles for these models are presented in Figures  \ref{fig:without01} and \ref{fig:without02}.

The comparison of the models without self-gravity R2 and R3 with the self-gravitating models N2b and N3, respectively (bottom panels in Figures \ref{fig:g0.6} and \ref{fig:MtoFInteg} and upper panels in Figures \ref{fig:g0.4} and \ref{fig:MtoF2Integ}) reveals  the importance of the inclusion of self-gravity, particularly  in  model R2 which does not develop a supercritical core contrary to its self-gravitating counterpart (model N2b). On the other hand,  model R1 (top panels in Figures \ref{fig:without01} and \ref{fig:without02}) is not much affected by the elimination of  self-gravity, as we can see when comparing with the self-gravitating model N1 (top panels of Figures \ref{fig:g0.6} and \ref{fig:MtoFInteg}). In this case the external gravitational potential due to the embedded stars is already very high and  dominates the cloud collapse making self-gravity non-negligible only in a very small radius close to the center. 

Model R1 is also comparable to the model D2 of Santos-Lima et al. (2010). Both have the same initial conditions and no self-gravity, but differ in the geometry, as the clouds in Santos-Lima et al. have initial cylindrical gravitational potential rather than spheric. The presence of a spherical gravitational field  yields a smaller efficiency in the magnetic flux transport, as we can see when comparing the top-middle diagram  of Figure \ref{fig:g0.6} with top-right panel of Figure 11 of Santos-Lima et al. (2010).  This is  because, while in the case of a spherical potential  all matter is pushed to a single central point, in the case of a cylindrical potential, gravity pushes the collapsing material to the central axis all along the cylinder and thus it is more effective to help the decoupling between the collapsing gas and the magnetic flux driven by the reconnection diffusion. 

\section{Discussion}
\label{Discussions}

We have performed 3D MHD simulations of the evolution of self-gravitating  cloud clumps with a central gravitational potential with spherical symmetry and  embedded in an initially uniform magnetic field. For comparison we have also considered a few models without self-gravity. The simulations were started with the cloud either in magneto-hydrostatic equilibrium (see model E1 of Table 1) or  out of it (all the other models of Table 1). We injected non-helical turbulence in each system with an rms velocity ($v_{turb}$) comparable both to the initial Alfvén speed and to the isothermal sound speed (see Table \ref{tab:non_equi}) and  then let it to evolve. All the tested models have an initial turbulent to magnetic energy ratio $> 1$, which is compatible with estimates from observations of the cold neutral interstellar medium (Heiles \& Troland 2005). We have imposed to all systems the same turbulent time decay (eq. \ref{eq:diff_time}) in order to compare the effects of the reconnection diffusion among them. Also for comparison, we evolved the same set of models with no injection of turbulence.
 
Most of the turbulent models here investigated evidence the transport of magnetic flux, i.e., the decoupling of the magnetic field from the denser, inner regions of the cloud clump due to the presence of MHD turbulence. The exceptions are model R3 without self-gravity (as demonstrated by Figures \ref{fig:without01} and \ref{fig:without02}) and the self-gravitating models N2c, N2d, and N4 (see Figures \ref{fig:g0.6b} and \ref{fig:MtoF3Integ}, and \ref{fig:g0.4} and \ref{fig:MtoF2Integ}, respectively). 

\subsection{Comparison with Santos-Lima et al. (2010) results}

Previous numerical studies of cylindrical cloud systems by Santos-Lima et al. (2010) had already evidenced the importance of the effects of turbulent magnetic reconnection to remove the magnetic flux excess from collapsing systems, as originally suggested by Lazarian (2005) based on the fact that magnetic reconnection must be fast in weakly MHD turbulent environments (Lazarian \& Vishniac 1999; Kowal et al. 2009; see also Appendix A). The present numerical study, which considers more realistic self-gravitating spherical clouds, confirms the results of this previous analysis, i.e., it shows that the presence of reconnection diffusion  is able to remove magnetic flux from the denser central regions to the edges of the cloud, therefore facilitating the gravitational collapse in most of the tested models, without considering ambipolar diffusion effects. This is assured  by the measured magnetic-flux-to-mass ratio in the core regions of the simulated clouds, which is quantified in our study by the average magnetic field-to-density ratio along the magnetic field lines (see middle diagrams of Figures \ref{fig:g0.6}, \ref{fig:g0.6b}, \ref{fig:equil}-top, \ref{fig:g0.4}, and \ref{fig:without01}). This ratio decreases with time in most of the turbulent models, while it remains constant, on average, in the non-turbulent counterpart models.  

As in Santos-Lima et al. (2010), we also find that, in general, an increase in the stellar gravitational potential (see e.g., models N1 and N2a of Figure \ref{fig:g0.6}), as well as a decrease in the initial magnetic field strength (or an increase in $\beta$; see models N2b of Figure \ref{fig:g0.6} and N2c and N2d of Figure \ref{fig:g0.6b}) favours the turbulent  reconnection transport of the magnetic flux and its decoupling from the dense collapsing gas. 

In the cases  when the turbulence is sub-Alfvénic (i.e., $v_{turb}/v_A <1$, as in models N2c, N2d, and N2e), the flux transport by RD is more difficult, as one should expect, since the turbulence at large scales is weak. In fact, for a cloud with an initially too large magnetic field, sub-Alfvénic turbulence will not be able to transport the magnetic flux to outside and the cloud may fail to build up a supercritical core, i.e., a core with a mass-to-magnetic flux ratio above the critical value that is required for gravity to overcome the magnetic forces (see for instance, models N2c and  N2d in Figures \ref{fig:g0.6b} and \ref{fig:MtoF3Integ} for which $\beta =$ 0.3 and 1.0, and $v_{turb}/v_A =$ 0.7 and 0.9, respectively). 

Nonetheless, even  sub-Alfvénic regimes of turbulence may allow the formation of critical cores (see results of Santos-Lima et al. 2010). This was the case of model N2e, which has  the same initial conditions of model N2c, except for a smaller total gravitational potential. The latter caused a delay of the gas collapse that gave time  for the sub-Alfvénic reconnection, which becomes stronger at smaller scales (Lazarian 2006),  to transport outward part of  the magnetic flux and allow the formation of a critical core (see Figure \ref{fig:MtoF3Integ}). 

The fact that both models N2c and N2e produce different outcomes is an indication that the overall process is sensitive to the initial conditions. Tested in higher resolution (512$^3$), both models confirmed the same trend, indicating that the system with smaller density has reached a minimum threshold for the onset of the gravitational instability which is overcome by the other forces.

The result above, can be understood in terms of the reconnection diffusion coefficient in sub-Alfvénic regimes. Although weak at large scales (see eq. \ref{eq:diffusivity} which shows that in this regime the RD is  smaller than in the  super-Afv\'enic regime by a factor $v_{turb}(v_{turb}/v_A)^3$), as the {\bf turbulence cascades the strength of the interactions increases and at a scale $l \simeq l_{inj}(v_{turb}/v_A)^2$ } it becomes stronger and therefore, more efficient to help with magnetic flux transport. As we will see in Section 4.2 below, the numerical diffusivity  is  smaller  than the effective reconnection diffusion operating at the relevant scales of these systems suggesting therefore, that the magnetic flux transport is dominated by RD\footnote{It should be noticed that, when N2e (or any other system) becomes unstable to the gravitational collapse, numerical viscosity will possibly also play a role at scales of the order or smaller than the sink region and influence the collapse runaway. This however will prevail only at very small scales well after the action of the turbulence at larger scales.}. In the next section we discuss the validity of this estimate for the reconnection diffusion coefficient. 

Another result of particular importance here is the fact that in the presence of a more realistic spherical gravitational field in the cloud, the magnetic flux transport is less efficient than in the presence of a cylindrical field (as investigated in Santos-Lima et al. 2010). This was already expected, since in a spherical potential  all gas is  pushed to a single central  point, while in the cylindrical field the gas is pushed to the central axis along the cylinder making   the decoupling of the gas from the magnetic flux, which is  driven by the reconnection diffusion, more effective in the latter. Therefore, we may conclude that the results of the previous study based on cylindrical turbulent clouds by Santos-Lima et al. (2010) have overestimated the flux transport by reconnection diffusion. For instance, for model R1 with spherical potential (top panels of Figure \ref{fig:g0.6}) the normalized magnetic flux-to-density ratio has decreased by a factor $0.11$ after 8 dynamical times, while the same model with cylindrical field in Santos-Lima et al. (2010) (model D2)  has decreased by a larger amount, $0.65$, at the same time interval.

\subsection{Comparison of turbulent magnetic reconnection diffusivity with resistivity effects and numerical diffusivity}
\label{ohmic}

In the study of cylindrical clouds by Santos-Lima et al. (2010), it has been found that the reconnection diffusivity is much larger than the numerical diffusivity both for initially trans and sub-Alf\'enic turbulent clouds. As in Santos-Lima et al. (2010), we can evaluate the effective reconnection diffusion coefficient of our simulated  models by comparing them with non-turbulent resistive models with enhanced uniform Ohmic resistivity. For instance, considering the same initial conditions as those of the trans-Alfvénic model N2b and the sub-Alfvénic models N2c, N2d and N2e, we performed several simulations of resistive non-turbulent models considering different values of enhanced Ohmic resistivity. Table \ref{tab:resistive} presents a set of resistive models N2br, N2cr, N2dr, and N2er whose initial conditions are the same as those of N2b, N2c, N2d and N2e, respectively.

\begin{figure}[!htb]
	\centering
		\includegraphics[width=7.5cm]{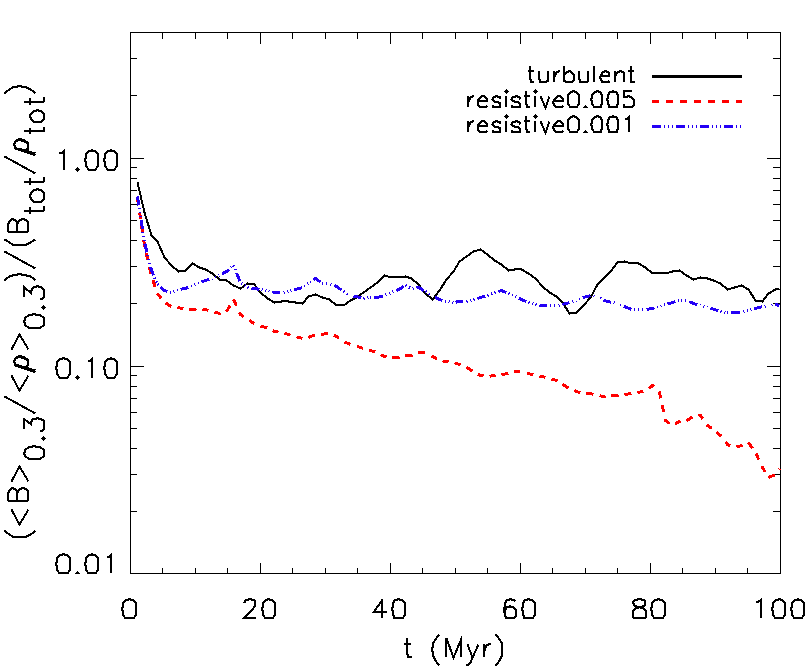}
	\caption{Temporal evolution of the average magnetic field-to-density ratio at the cloud core region within a radius $r_c =$ 0.3 pc normalized by the average value over the entire cloud, for the turbulent N2e model (black, solid line), the resistive model N2er1 with $\eta_{Ohm} =0.001$ c.u. (blue, dot-dashed line); and the resitive model N2er2  with $\eta_{Ohm} =0.005$ c.u. (red, dotted line). We note that the turbulent model N2e is comparable to the resistive model N2er1.}
	\label{fig:ueta}
\end{figure}

We have found that resistive models with  $\eta_{Ohm} \gtrsim 0.002$ c.u. are the ones which best reproduce model N2b, thus we can conclude that the effective turbulent resistivity of model N2b is $\eta_{turb} \gtrsim 0.002$ c.u., or  $\gtrsim 6 \times 10^{20}$ cm$^2$ s$^{-1}$ in physical units\footnote{We note that in the case of this model (N2b), after the full development of the turbulence in the cloud, the gravitational core collapse is very fast making it difficult a straightforward comparison with resistive models with similar initial conditions. This explains why in this case we could only estimate a lower limit for the effective turbulent diffusion coefficient.}. A similar procedure for models N2c, N2d and N2e allowed us to estimate also approximate values for their turbulent diffusion coefficients, which are presented in Table \ref{tab:eta_turb}. Figure \ref{fig:ueta} illustrates this comparison between turbulent and resistive models for N2e and its resistive counterparts N2er1 and N2er2. Table \ref{tab:eta_turb} also shows the corresponding estimates for the ratio $\eta_{turb}/l_{inj} v_{turb}$ for these models, calculated in the core region of the cloud at the final time step\footnote{This value  changes only by a  factor two, at most, with regard to its  initial value.}. According to eq. \ref{eq:diffusivity}, we should expect this ratio to be  smaller than 1 for sub-Alfvénic turbulence  (i.e., of the order  of the Alfvénic Mach number to the third power) as indicated in the last column. We notice also that model N2b, which started with trans-Alfv\'enic turbulence, ended up with sub-Alfv\'enic turbulence in the core too due to the  increase of the magnetic field (and thus of the Alfv\'en speed) with the collapse. Therefore, the obtained {\bf values are roughly consistent with the theoretical predictions, although} further theoretical developments are still required in order to obtain a more precise value of the reconnection diffusion coefficient within the MHD theory of turbulence (see Lazarian 2011 and references therein).
 
We must remark that the relevant scales for RD to operate are the turbulent scales, from the injection scale ({\bf of the order of the cloud size)  to the dissipation scale, i.e., within the inertial range scales of the turbulence, which are generally much larger than the numerical diffusion scale.} In general lines, from the comparison between the effective values of the RD coefficients and the numerical diffusion in our models, we find that the first is larger than $\eta_{Num}$  (see also Santos-Lima et al. 2010), especially in the case of initially super and trans-Alfv\'enic clouds. In our numerical simulations with a resolution of $256^3$, the numerical diffusion scale is approximately of 8 cells, corresponding to a numerical diffusivity in these models roughly around $\eta_{Num} \sim 10^{-4}$ c.u or $\eta_{Num} \sim 3 \times 10^{19}$ cm$^{2}$ s$^{-1}$ \footnote{This value was estimated from the comparison of numerical simulations of  pure resistive models  with laminar ones.}. For runs performed with a $512^3$ resolution, the numerical diffusivity decreases by a factor two, or $\eta_{Num} \sim 5 \times 10^{-5}$ c.u. $\sim 0.15 \times 10^{19}$ cm$^{2}$ s$^{-1}$ which is much smaller than the effective RD diffusivity found for the models (see Table \ref{tab:eta_turb}), therefore, clearly indicating that the magnetic flux removal we see in the models is due to turbulent reconnection diffusion and not to numerical dissipation.

\begin{table}[!htb]
\caption{Parameters for resistive models with same initial conditions 
of the initially trans-Alfvénic model N2b and the sub-Alfvénic models N2c, N2d and N2e}
 \vspace{1.5ex}
 \centering
 \begin{tabular}{l c c c}
	\hline \hline\\
       [-1.0ex]
	Model 	&  n (cm$^{-3}$)	&	$\beta$    &    $\eta_{Ohm} $ ($3\times 10^{23}$ cm$^2$s$^{-1}$)	\\
      [0.5ex]
	\hline \\
      [-1.2ex]
    N2br1	&	 $90.0$     &	 $3.0$     & 	  $0.001$	\\
	N2br2	&	 $90.0$     &	 $3.0$     & 	  $0.005$	\\
	N2br3	&	 $90.0$     &	 $3.0$     & 	  $0.01$	\\
	N2br4	&	 $90.0$     &	 $3.0$     & 	  $0.05$	\\
    N2br5	&	 $90.0$     &	 $3.0$     & 	  $0.1$	\\
	N2cr1	&	 $90.0$     &	 $1.0$     & 	  $0.001$	\\
	N2cr2	&	 $90.0$     &	 $1.0$     & 	  $0.005$	\\
	N2cr3	&	 $90.0$     &	 $1.0$     & 	  $0.01$	\\
	N2cr4	&	 $90.0$     &	 $1.0$     & 	  $0.05$	\\
	N2cr5	&	 $90.0$     &	 $1.0$     & 	  $0.1$	\\
	N2dr1	&	 $80.0$     &	 $1.0$     & 	  $0.001$	\\
	N2dr2	&	 $80.0$     &	 $1.0$     & 	  $0.005$	\\
	N2dr3	&	 $80.0$     &	 $1.0$     & 	  $0.01$	\\
	N2er1	&	 $90.0$     &	 $0.3$     & 	  $0.001$	\\
	N2er2	&	 $90.0$     &	 $0.3$     & 	  $0.005$	\\	
 	  [0.5ex] \hline
 \end{tabular}
 \label{tab:resistive}
\end{table}

\begin{table}[!htb]
 \begin{center}
\caption{Diffusivity coefficient for the initially trans-Alfvénic model N2b and the sub-Alfvénic models N2c, N2d and N2e}
 \vspace{1.5ex}
 \centering
 \begin{tabular}{l c c c}
	\hline \hline\\
       [-1.0ex]
	Model 	& $\eta_{turb}$ ($3 \times 10^{23}$ cm$^2$s$^{-1}$) & $\eta_{turb}/v_{turb}l_{inj}$	& $(v_{turb}/v_A)^3$  \\
      [0.5ex]
	\hline \\
      [-1.2ex]
	N2b	   	&				 $\gtrsim 0.002$    				  & 		  $> 0.0028$		    &	 $\sim0.073$	  \\
	N2c		&				 $\sim 0.001$     				  & 		  $\sim 0.0024$		&	 $\sim0.229$	  \\
	N2d		&				 $\sim 0.001$     				  & 		  $\sim 0.0016$		&	 $\sim0.096$	  \\
	N2e		&				 $\sim 0.001$     				  & 		  $\sim 0.0027$		&	 $\sim0.287$	  \\
 	  [0.5ex] \hline
 \end{tabular}
 \label{tab:eta_turb}
 \end{center}
\end{table}

As stressed in Section \ref{intro}, these conclusions are also supported by the previous numerical studies of Santos-Lima et al. (2010, 2012, 2013) and besides, indicate the convergence of the results already at a $256^3$ resolution. On the other hand, they clearly indicate a smaller efficiency of the RD at sub-Alfv\'enic regimes of turbulence, as predicted by the theory, which in some cases may even require higher resolution simulations in order to ensure that the effective RD coefficient is larger than the numerical diffusivity.

Another point that can be noticed is the fact that the calculations here were performed under an one-fluid approximation and this does not allow for the action of ambipolar diffusion. Nonetheless, only for completeness we can compare the RD coefficients estimated above with the expected ambipolar diffusion coefficient for the corresponding physical conditions of the cloud clumps here investigated. This is of the order of $\eta_{AD} \simeq v_A^2 t_{n,i} \sim 10^{15}$ cm$^{2}$ s$^{-1}$, where  $t_{n,i} \sim 4.8\times 10^8/n_i$ is the mean time of neutral-ion collisions and $n_i$ is the ion number density, which is also much smaller than the RD coefficients evaluated above.

\subsection{Effects of self-gravity upon the magnetic flux transport by reconnection diffusion}

The comparison of  models without self-gravity  with  self-gravitating models  have revealed that self-gravity can also significantly help the decoupling between gas and magnetic flux due to reconnection diffusion, particularly in the late stages of the cloud collapse. A critical example is the self-gravitating model N2b (Figures \ref{fig:g0.6} and \ref{fig:MtoFInteg}) in which reconnection diffusion causes the build up of a supercritical core, while its counterpart without self-gravity, model R2 (Figures \ref{fig:without01}), is unable to develop a supercritical core.

An increase of the  self-gravity (which  is provided by an increase in the initial gas density of the cloud) improves the turbulent transport of the magnetic flux. If the gas density  in the cloud is large enough ($n_0 > 50$ cm$^{-3}$), its effect seems to be more important than that of the stellar gravitational potential (for $M_{\star} \sim 41$ M$_{\odot}$)  to help the decoupling between the gas and the magnetic field (see models N2b and N3 in Figures \ref{fig:g0.6} and \ref{fig:MtoFInteg} and \ref{fig:g0.4} and \ref{fig:MtoF3Integ}). However, for a given strength of turbulence, if one increases the initial density or the stellar gravitational field indefinitely, then eventually  the total gravitational potential will become so high that it will neutralize the ability of the reconnection diffusion to decouple the magnetic flux from the dense gas and thus, no efficient transport of magnetic flux will occur to outside of the cloud core, as most of it will be dragged by the infalling gas. We have seen this effect, for instance, when  increasing the initial cloud density in  model N2e (which has initial $\beta =1.0$,  $M_\star \sim 41$ M$_{\odot}$, and $n_0=80$ cm$^{-3}$)  to $n_0=90$ cm$^{-3}$ in model N2c (see Figures \ref{fig:g0.6b} and \ref{fig:MtoF3Integ}). While the first model evidences some magnetic flux transport and develops a nearly critical core, the second one fails completely. Similarly, for model N2b (which has $\beta = 3.0$,  $M_\star \sim 41$ M$_{\odot}$, and $n_0=90$ cm$^{-3}$), transonic, trans-Alfvénic turbulence allows the formation of a supercritical core, as indicated in Figures \ref{fig:g0.6} and \ref{fig:MtoFInteg}. However, if one increases its initial gas density to $100$ cm$^{-3}$, the total gravitational potential becomes so large and makes the collapse so fast that the turbulent magnetic reconnection becomes ineffective to decouple the magnetic flux from the dense core. On the other hand, if one reduces the stellar mass of model N2b to M$_\star \sim 27$ M$_{\odot}$ and increases the gas density to $100$ cm$^{-3}$ as in model N3 (Figures \ref{fig:g0.4} and \ref{fig:MtoF2Integ}), so that the total mass is nearly the same as in model N2b, then some flux transport is evidenced and a supercritical core develops, but a further reduction of the total mass (as in N4) again prevents the formation of a critical core, because in this case the infall becomes so slow that the turbulence actually helps to spread out the core material.

\subsection{Effects of the cloud initial conditions}

All the results above were found for cloud clumps which had initial uniform density and were  out of magneto-hydrostatic equilibrium when turbulence was injected. We have also tested a model starting in magneto-hydrostatic equilibrium having a stratified density,  with  the  central density and the remaining initial conditions as in our reference model (N2b) (see model E1 in Figure \ref{fig:equil}). As N2b, model E1 also undergoes  an efficient outward magnetic flux transport due to reconnection.  However, the much smaller total mass of E1 due to the cloud stratification  (with a fraction 0.43 of the total mass of N2b model)  prevents  its collapse to form  a supercritical core within the evolved time interval.

The results above indicate that the formation of a supercritical core is regulated by a complex interplay  between gravity, self-gravity, the magnetic field strength and  nearly transonic, trans-Alfvénic turbulence.  Although we have found that reconnection diffusion is very efficient to remove magnetic flux from most of the collapsing core clump models tested here, only a few were succeeded to develop nearly critical or supercritical cores (see models R1, N1, N2b, N2e, and N3) which may be able to collapse and form stars. In other words, for the cloud conditions investigated here, the formation of supercritical cores is restricted to   a limited  range of  parameters, as one actually should expect from observations that predict a low efficiency of star formation (see e.g. Mac Low \& Klessen 2004; Le\~ao et al. 2009; Vazquez-Semadeni et al. 2011). 

To summarize, our results suggest that in the presence of nearly transonic and trans-Alfv\'enic turbulence, flux transport by RD will allow initially sub-critical clouds to become nearly critical or supercritical. This condition will be fulfilled for cloud clumps with initial values of $\beta \sim 1$ to 3, cloud densities $50 < n_0 < 100$ cm$^{-3}$ when considering stellar masses $M_{\star}\sim 41$M$_{\odot}$, and  densities $100 < n_0 < 140$ cm$^{-3}$ when considering stellar masses $M_{\star}\sim 27$M$_{\odot}$, implying total cloud clump masses $M_{tot}\lesssim 120$M$_{\odot}$. For smaller densities the clouds are fragmented  by the injected  turbulent power and no core is built up. For higher densities, the effects of  self-gravity are so strong that the core collapses, dragging most of the magnetic field, so that no significant magnetic flux transport is detected. Of the 9 self-gravitating models here investigated, 4 formed marginally critical or supercritical cores (N1, N2b, N2e and N3) and 2 subcritical cores (N2a and E1), all of which evidenced turbulent  magnetic flux transport. The 3 remaining models (N2c, N2d and N4) did not evidence any magnetic flux transport by reconnection diffusion (either because the initial cloud had too strong magnetic field or too strong turbulence, as described in Section 3). Table \ref{tab:ratios} lists the final conditions of the built-up cores of all simulated models.

\subsection{Effects of decaying turbulence}

The resulting dense cores built up  in the simulations are characterized by turbulent energies of the same order or even larger  than the magnetic energy. This is in contrast with observed dense cores which have little internal turbulent energy left as compared to their gravitational and magnetic energies (see Bergin \& Tafalla 2007; Pineda et al. 2010). In fact, in our  simulations, for simplicity, the turbulence was continuously injected, making it nearly constant (over the system) with time. Thus, after magnetic flux was partially removed from the core by RD, the resulting turbulent to magnetic energy ratio increased in most of the cases with respect to the initial values of the diffuse cloud clumps. Nevertheless, we have found  that once the magnetic flux is partially removed allowing the formation of a critical or supercritical core, the turbulence has no more significant dynamical effect upon its evolution which is then dominated by gravity. For instance, for models N1 and N2b (see Figure \ref{fig:g0.6} and \ref{fig:g0.6b}) we have found that just a few dozen Myr after the full development of the turbulence in the system (around $\sim$20-40 Myr in Figure \ref{fig:g0.6}), critical and supercritical cores have already developed in the central regions of the clouds, respectively, so that turbulence had no more relevant effect upon them from that time on.

In order to exemplify this issue more quantitatively, Figure \ref{fig:N2b_off} in the  Appendix  (see also Table \ref{tab:ratios}) shows the results for a model with the same initial conditions of model N2b, but where the turbulence was  turned off after 50 Myr. We find that thanks to the decaying of the turbulence, the cloud clump develops a supercritical core much more efficiently and earlier (see also  Figure 2 for comparison). In other words, while the turbulence helps the initial collapse by removing the excess of magnetic flux through  RD action, later its continuous injection, as in N2b model (Figure 2),  can make the infall more difficult, mostly in the surrounding envelope. The left and middle panels of Figure \ref{fig:N2b_off}, which show edge-on and  face-on cuts, respectively, of the central regions, reveal the formation of a supercritical core (with a mean mass-to-flux ratio $\sim 30$\footnote{We note that after the turbulence is turned off the core still pulsates at the same time that it collapses. The estimated mass-to-flux ratio has been averaged over these oscillations.}) with a disk-like structure surrounding it. Besides, we also find that the resulting turbulent to magnetic energy ratio in the core in this case is of the order of 1.68, which is consistent with the observations. This issue of the  turbulence decay will be further explored in a forthcoming work. 

\begin{table*}[!htb]\footnotesize
 \begin{center}
\caption{Final quantities obtained for the built up cores and envelopes. Central density $n_c$ and magnetic field in z-direction $B_c$, mass $M_{c}$, magnetic flux $\Phi_{c}$, and maximum mass-to-magnetic flux ratio relative to the critical value $\mu_{crit,c}$ for the built up cloud cores ($r_c \leq 0.3$ pc). Mass $M_{e}$ and magnetic flux $\Phi_{e}$ for the formed cloud envelopes;  mass-to-magnetic flux ratios between the cloud core and the envelope, $R$, mass-to-flux ratios between the cloud core and the entire cloud (core$+$envelope) cloud $R^{'}$, at the final time step ($t\sim100$ Myrs). Models N1, N2b, N2e, N3 and R1 developed supercritical or marginally critical cores. All models but N2c, N2d, and N4, evidenced magnetic flux transport by reconnection diffusion (RD). }
 \vspace{1.5ex}
 \centering
 \begin{tabular}{l c c c c c c c c c c}
	\hline \hline\\
       [-1.0ex]
	  Model  & $n_c$(cm$^{-3}$) & $B_c$($\mu$G) &  $M_{c}$(M$_{\odot}$)  &  $\Phi_{c}$($10^{31}$G.cm$^{2}$)  & $\mu_{crit,c}$ &  $M_{e}$(M$_{\odot})$  & $\Phi_{e}$($10^{31}$G.cm$^{2}$) & $R$ &  $R^{'}$  \\
      [0.5ex]
	\hline \\
      [-1.2ex]
      R1      &  $3.8\times10^{3}$  &  $1.92$  &      $0.7$    &     $0.20$     & $1.32$  &     $7.7$    &   $3.54$    &  $1.52$  &  $1.47$   \\
      R2      &  $4.2\times10^{3}$  &  $4.04$  &      $2.0$    &     $0.62$     & $0.96$  &    $73.7$    &   $10.6$    &  $0.46$  &  $0.48$   \\
      R3      &      $994.7$        &  $3.32$  &      $1.1$    &     $0.43$     & $0.73$  &    $83.0$    &   $11.2$    &  $0.33$  &  $0.34$   \\
      [0.5ex]
	\hline
	  N1     & $3.5\times10^{3}$   &  $2.60$  &      $0.6$    &     $0.23$     &  $1.2$  &     $7.8$    &   $0.35$    &  $1.29$  &  $1.27$   \\
	  N2a    &      $505.3$        &  $1.44$  &      $0.2$    &     $0.15$     &  $0.6$  &     $8.2$    &   $3.73$    &  $0.73$  &  $0.74$   \\
	  N2b    & $1.1\times10^{7}$   &  $89.6$  &     $13.2$    &     $0.84$     & $40.7$  &    $62.5$    &   $11.1$    &  $2.80$  &  $2.49$   \\
	  N2c    & $5.6\times10^{3}$   &  $8.75$  &      $1.4$    &     $0.68$     &  $0.7$  &    $74.3$    &   $19.4$    &  $0.53$  &  $0.54$   \\
	  N2d    & $2.9\times10^{3}$   &  $4.14$  &      $1.0$    &     $1.43$     &  $0.2$  &    $74.7$    &   $35.4$    &  $0.34$  &  $0.35$   \\
	  N2e    & $1.3\times10^{3}$   &  $9.00$  &      $1.2$    &     $0.92$     &  $1.5$  &    $66.1$    &   $18.2$    &  $0.35$  &  $0.36$   \\
	  N3     & $2.3\times10^{3}$   &  $3.79$  &      $1.7$    &     $0.31$     &  $2.0$  &    $82.4$    &   $11.8$    &  $0.79$  &  $0.79$   \\
	  N4     & $1.3\times10^{3}$   &  $3.29$  &      $1.4$    &     $0.51$     &  $0.8$  &    $82.7$    &   $11.8$    &  $0.38$  &  $0.39$   \\
       E1     &      $111.6$        &  $0.85$  &      $0.1$    &     $0.11$     &  $0.3$  &     $2.2$    &   $1.93$    &  $0.39$  &  $0.41$   \\
      N2b-off & $3.2\times10^{7}$   &  $92.2$  &     $35.0$    &     $0.81$     & $29.0$  &    $40.7$    &   $10.6$    &  $4.76$  &  $3.29$   \\
 	  [0.5ex] \hline
 \end{tabular}
 \label{tab:ratios}
 \end{center}
\end{table*}

\subsection{Comparison of our results with observations}
\label{observations}

A comparison of the theoretical expectations of the reconnection diffusion with the observational data was recently presented in Lazarian, Esquivel \& Crutcher (2012, henceforth LEC12). It showed reasonable agreement between the theory and observations. However, to provide detailed quantitative comparisons one has to perform numerical modelling as in this paper. In particular, LEC12 compared the rate of reconnection diffusion with the rate of gravitational collapse and established that for given velocity dispersions known from observations there is a range of densities which are not expected to produce an increase of magnetic field intensity with the increase of gaseous density. This agrees well with observations, but contradicts the flux freezing idea\footnote{Flux freezing should be well satisfied in highly ionized diffuse gas where ambipolar diffusion is negligible.}. LEC12 estimated that for the densities of the order of $10^{4}$ cm$^{-3}$ the gravitational collapse should prevail as RD process becomes too slow to remove the magnetic flux on the free fall time scale. This conclusion was found to be in good agreement with observations.

\begin{figure*}[!htb]
	\centering
		\includegraphics[width=12cm]{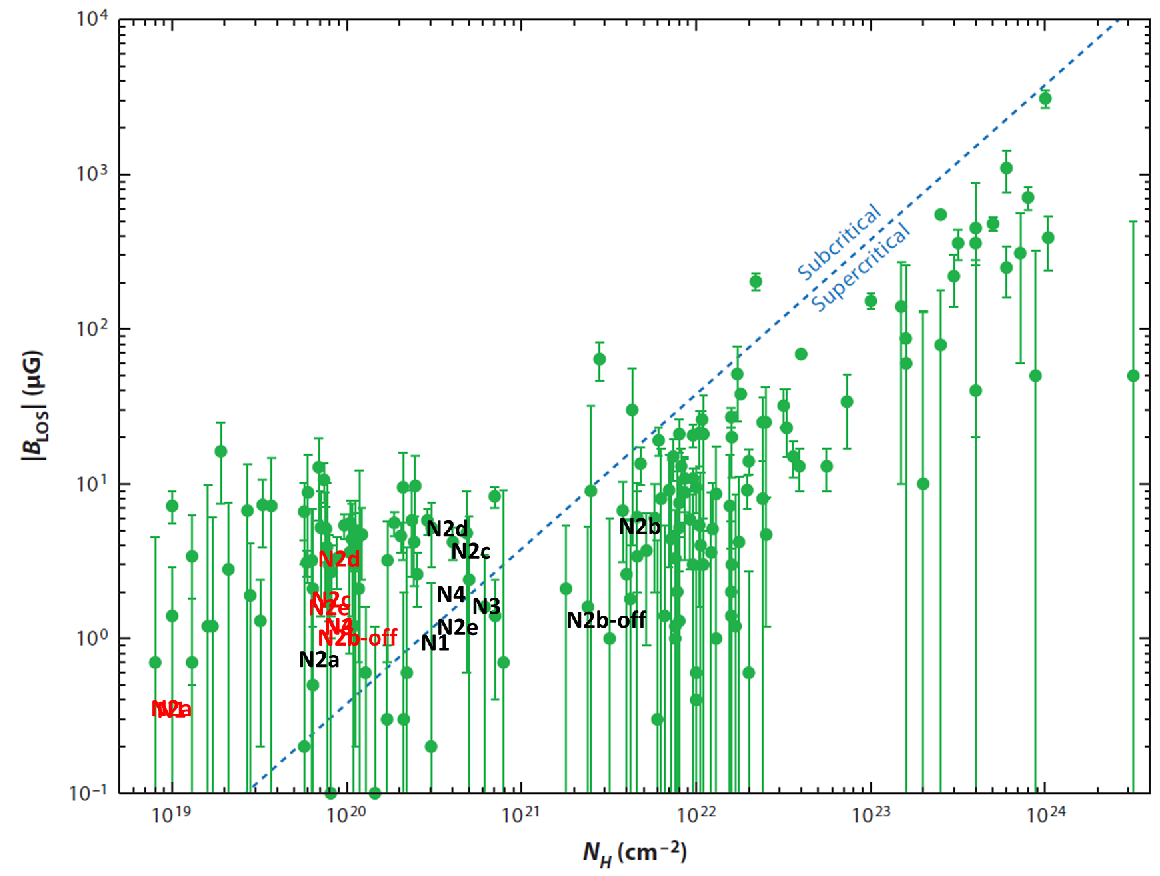}
	\caption{The simulated self-gravitating models are superposed to the observed interstellar cores diagram of magnetic field versus column density (Crutcher 2012).  Red labels the initial, and black the final states of the cores of the models of Table 1 (with  radius $r\sim 0.3$pc and magnetic fields  averaged over the entire core). Model N2b-off has been also included. The non self-gravitating models have not been plotted, but the final state of model R1 coincides with model N1.  The green dots correspond to HI, OH, and CN Zeemann measurements of the magnitude of the magnetic fields along the line of sight, ($B_{los}$ versus $n_H$) for the observed cores. The straight blue line separates the subcritical region on the left of the diagram from the supercritical one on the right side of the diagram.}
	\label{fig:testes}
\end{figure*}

With our present simulations we cannot provide the detailed testing of the theoretical expectations, but we still can probe some regimes of RD. Figure \ref{fig:testes} depicts the simulated models with self-gravity at their initial and final states (labelled with red and black colors, respectively) superposed to the observed diagram of magnetic field  strength versus column density for interstellar cores \citep{Crutcher2012}. The dashed straight line in this diagram separates the subcritical (on the left) from the supercritical region (on the right). This diagram clearly shows  that from the 12 simulated clumps of Table 1 only five develop critical or supercritical cores, as stressed above\footnote{The models R with no  self-gravitaty  have not been included in the diagram, but model R1 coincides with model N1.  Model N2b-off has been also included in the diagram for completeness.}. Apparently, for the parameters chosen we do not see substantial increase in the magnetic field in the evolved supercritical cores. According to LEC12, this corresponds to the regime of fast reconnection diffusion and agrees well with our expectations.

Another puzzle (also addressed in LEC12) is related to the recent Zeemann measurements of dark cloud cores and envelopes by Crutcher et al. (2009, 2010; see also Troland \& Crutcher 2008). The ambipolar diffusion (AD) is expected to be slow in the envelopes due to the high ionization and therefore, magnetic flux escaping from the core should be trapped in the envelope if the AD is the dominant process of flux diffusion. Quite surprising, the results in Crutcher et al. (2009, 2010) were the opposite. Namely, they obtained the mass-to-magnetic flux ratio between the cloud core and the envelope, $R=(M_c/\Phi_c)/(M_e/\Phi_e)$, as well as  the mass-to-flux ratio between the core and the entire (core$+$envelope) cloud, $R'=(M_c/\Phi_c)/(M_{c+e}/\Phi_{c+e})$, and  found that these ratios are less than unity for four observed clouds (B1, B217-2, L1448CO and L1544). The requirement of $R'>1$ is a necessary one for the AD model, while the RD process can enable the opposite ratio. In fact, LEC12 pointed out that the stronger and larger scale turbulence in the envelope is bound to remove magnetic flux faster from the envelope compared to the cloud core  (a result also verified in Santos-Lima et al. 2010). 

The conditions of the cloud core-envelopes  investigated by Crutcher et al. (2009, 2010) are somewhat distinct from those in our numerical simulations. Nevertheless, we can make at least qualitative comparisons with their results. For instance, although denser and more magnetized, their cores are all subcritical or only marginally critical (i.e., they have  mass-to-magnetic flux ratios relative  to the critical value in the range $\mu_{crit} = 0.45 - 1.15$). The cores built up in our models which evidenced  turbulent flux transport have final average mass-to-magnetic flux ratios $\mu_{crit} = 0.15 - 5.25$. Most of these cores have ratios $R$ and $R'$ which are consistent with the inferred ones by Crutcher et al. (2009), with the exception of the cores built up in models N1 and Nb2 (see Table \ref{tab:ratios}, where these ratios are presented for the final snapshots of all our models). Indeed, observationally, it is known that the level of turbulence drops in the cores. As expected theoretically (Lazarian 2006) reconnection diffusion slows down with the decrease of the turbulent velocity (see numerics in Santos-Lima et al. 2010), so that  we should expect a slower transport of magnetic flux from the core as compared to the envelope. This behaviour is actually detected in the cores formed in the present study and  provides more justification for the considerations in LEC12.

In the particular case of model N2b, the computed ratios $R$ and $R'$ are smaller than unity until the core becomes highly supercritical, from this point on $R$ and $R'$ become larger than unity due to the runaway increase of the mass-to-flux ratio of the collapsing core (maximum $\mu_{crit,c} \sim 40.7$, as shown in Table \ref{tab:ratios} and Figure \ref{fig:MtoFInteg}). {\bf At the same time, in the case of model N1, the} ratios $R$ and $R^{'}$ are also smaller than 1 for several time-steps. After 68 Myrs, the very rapid increase of the central density caused by the high central gravitational potential in this model, makes $R^{'}$ and $R$ to increase to values larger than 1 (the same situation applies to the non-self-gravitating model R1). Therefore, the three cores end up with $R$ and $R^{'}>1 $ because they are already collapsing to form proto-stars, N2b due to the dominance of self-gravity and N1 (and R1) due to the strong central potential. 

The other supercritical cores (N2e  and N3) still have values of $R$ and $R^{'} < 1$ at the final snapshot, therefore comparable to those obtained for the observed cores by Crutcher et al. However, if integrated for longer time, these collapsing cores will possibly change to the regime where $R$ and $R^{'}>1 $, as in N2b and N1 cases.

Also, it should be remarked that the ratios $R$ and $R^{'}$ have been computed with the  values of the mass to flux ratios integrated over the radius. Therefore, the larger the core radius the larger its integrated value. This is also the case for the radii encompassing the envelopes around the cores. This helps to understand why even for supercritical cores  $R$ and $R^{'} $ may be larger than unit (probably depending on the collapsing stage). We have found that  when computing  the differential mass to flux ratio $dM/d\Phi$ as a function of the radius (rather than the integrated one), the corresponding values in the envelopes are smaller than those in the cores for all critical and supercritical models (see examples in Figure \ref{fig:dMtodPhi} in Appendix).

It should be noted also that the inferred values for $R$ and $R'$ from the observations are subject to significant uncertainties due to measurement limitations and simplified assumptions (see Mouschovias \& Tassis 2010). Therefore, while encouraging, the results from the comparisons above should be viewed with caution. Besides, they call for further and more precise observational estimates of these ratios.

\begin{figure*}[!htb]
  \centering	
	\includegraphics[width=0.8\linewidth]{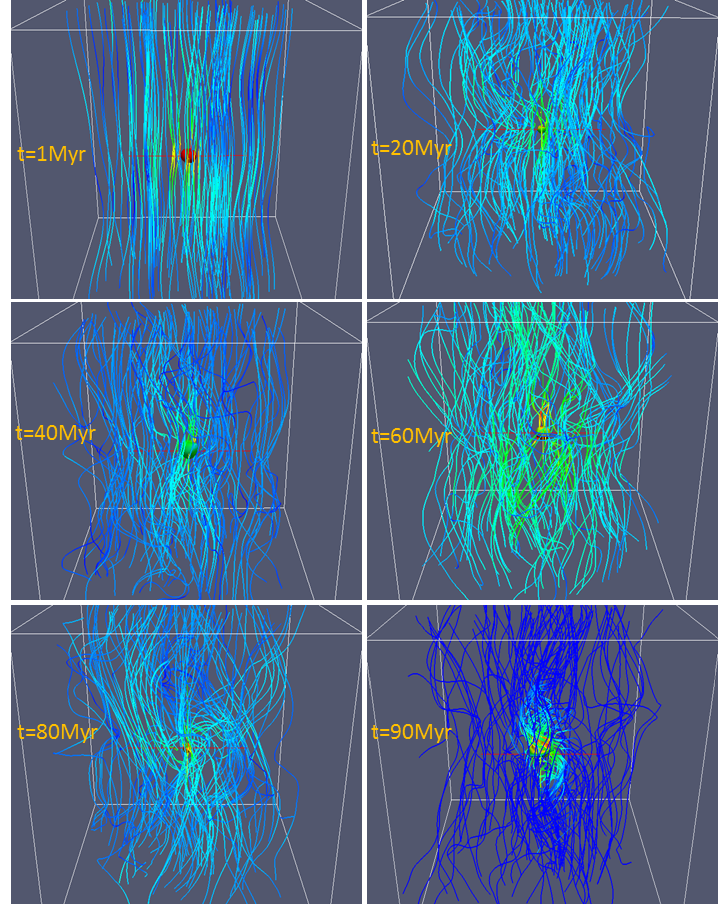}
  \caption{Time evolution of the magnetic field lines of the turbulent model N2b from  $t=$ 1 to 90 Myr (one-hundred equally spaced magnetic field lines are plotted within a radius of $\sim 1.3$ pc.). We clearly see how turbulence diffusively transports the magnetic field lines as time evolves. At t= 90 Myr the lines that were not transported to the outskirts of the cloud are advected to the center by the collapsing gas causing a local enhancement in the magnetic field intensity.  The color scale indicates the intensity of the magnetic field with red representing the maximum and blue the minimum values.}
  \label{fig:N2b_pw1_t}
\end{figure*}

An additional advantage of numerical simulations is that they can reveal the structure of magnetic field. This structure is becoming available through different types of polarization observations, both in emission and extinction\footnote{The polarization is believed to arise from non-spherical dust grains aligned in the magnetic field (see e.g., Lazarian 2007, Anderson 2013 for  reviews).}. Figure \ref{fig:N2b_pw1_t} illustrates the transport of the magnetic field lines from the collapsing core to the surrounding envelope for model N2b until the start of the core collapse around 90 Myr when then, the portion of the magnetic flux that was not diffusively transported to the surrounding envelope is advected to the center by the collapsing gas causing a large increase of the magnetic field intensity in this region.

\begin{figure*}[!htb]
  \centering	
    	\includegraphics[width=0.8\linewidth]{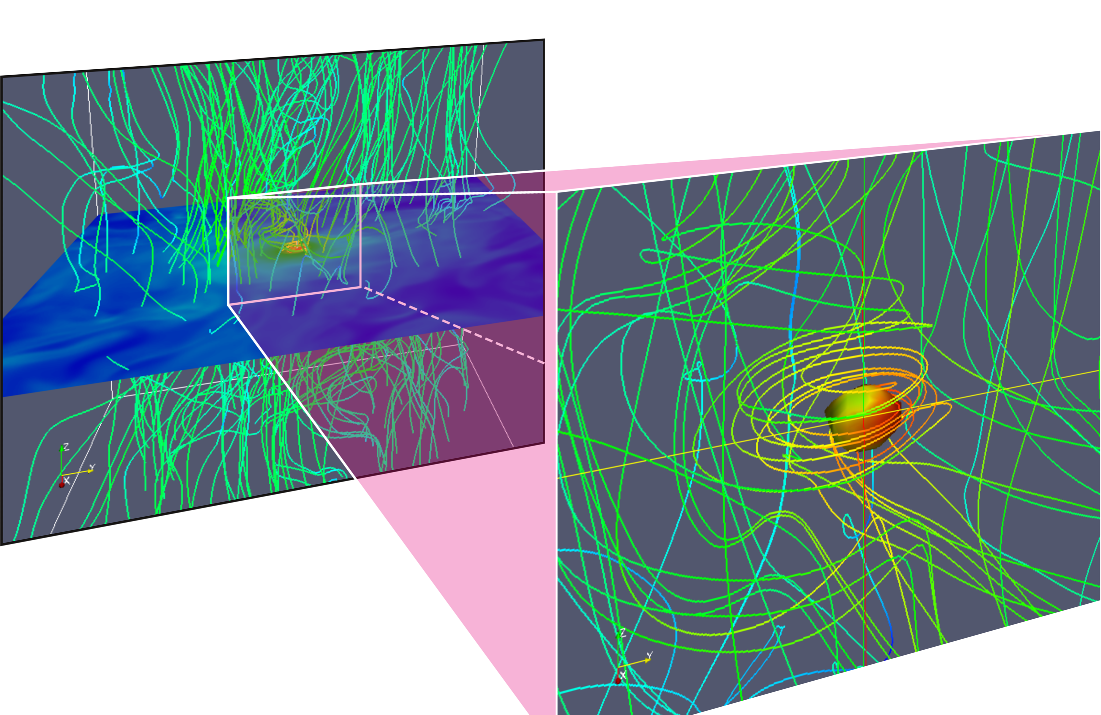}
  \caption{Left panel: magnetic field line distribution and the equatorial cut of the logarithmic density distribution for the collapsing cloud model N2b around 100 Myr. The central core has 100 equally spaced magnetic field lines are plotted within a radius of $\sim 1.3$ pc. Right panel: zoom out of the central region of the same system showing the magnetic field line distribution and the surface plot of the logarithmic density distribution for the formed supercritical core. The density in the plotted surface is $n=5000$ cm$^{-3}$ and corresponds to the core region with radius $\sim 0.12$ pc $\sim 2.5\times 10^{4}$ AU. Note the dominating helical magnetic field structure that develops around the core.}
  \label{fig:N2b_final}
\end{figure*}

Figure \ref{fig:N2b_final} depicts the same turbulent model (N2b) when the supercritical core has formed. The right panel of this figure highlights the supercritical core region. We clearly note that the magnetic field geometry is much more organized there and predominantly helical, although elongated magnetic field lines are also present. (A similar geometry has been detected also in the case when turbulence was switched off earlier in the evolution of the core, as in model N2b-off.) This is fully consistent with recent sub-millimetric polarization observations of several cores by Tang et al. 2009; Girart et al. 2006; Girart et al. 2009, Hull et al. 2012; and Lai et al. 2002, which evidence more organized laminar distributions of the magnetic field and, in most cases with a predominant component aligned with the oblate core, as in Figure \ref{fig:N2b_final}.

\section{Concluding remarks}
\label{Conclusions}

The mechanism here discussed of turbulent reconnection diffusion (RD) may present the last missing piece for constructing the new paradigm of star formation where turbulence and turbulent feedback play a central role. The mechanism is based on solid theoretical foundations routed both in our understanding of magnetic reconnection in turbulent media (LV99) and in connection to it, violation of the magnetic flux freezing in turbulent fluids (Eyink 2010; Eyink et al. 2011). Due to this fact we can demonstrate that the substantial differences between the Ohmic-type diffusivities  of the numerical simulations and those of the real interstellar media marginally affect our numerical modelling,  as far as magnetic diffusion is concerned. Indeed, magnetic reconnection and reconnection diffusion are determined in both cases by large scale turbulence rather than by the physical processes on micro-scales which are actually impossible to resolve with current computing.

The RD process should not be confused with an increase of magnetic resistivity. While the resistivity in our simulations does not depend on the level of turbulence in the system, both the  reconnection rate and the diffusion of matter and magnetic fields are determined by the intensity and the injection scale of turbulence. We should  stress that the RD mechanism may have important implications also in turbulent dynamo processes (see de Gouveia Dal Pino 2012, 2013). This possibility will be investigated numerically elsewhere. 

The present numerical study together with the earlier one by Santos-Lima et al. (2010) have investigated the RD mechanism focusing on the early stages of star formation. Recently, Santos-Lima, de Gouveia Dal Pino \& Lazarian (2012, 2013) have studied this mechanism  in the late stages, during the formation of  protostellar disks. They have shown, {\bf by means of  3D MHD simulations, that turbulent RD is also  able to transport magnetic flux from the disk progenitor at time scales which are  consistent with the core} collapse. 
A rotationally supported disk with a nearly Keplerian profile was shown to build up  around the protostar in only a few  $10^4$yr, as required by the observations. Therefore, taken together these studies have tested the new paradigm of magnetic flux removal from molecular clouds by turbulent reconnection diffusion. The RD efficiency especially when considering initially trans or super-Alfv\'enic clouds, calls for revaluation of the relative role played by ambipolar diffusion (and other transport mechanisms) in the processes of star and planet formation. Recent studies undertaken by Myers et a. (2013) are also consistent with our results.

Finally, we should remark that in this study, we have focused on the evolution of isothermal self-gravitating clouds with embedded stars providing an external gravitational potential.  As a matter of fact, some of the mentioned observed dark cloud cores contain  stars embedded in them (e.g., B1 and L1448CO clouds; Bachiller, Menten, \& del Rio-Alvarez 1990; Volgenau et al 2006). In forthcoming work, we will explore the effects of the magnetic flux transport by RD in the evolution of initially starless clouds in order to assess the effects of self-gravity only upon the transport, without considering an external field. Also, the isothermal approximation  assumed here mimics the effects of an efficient radiative cooling of the gas. However, in  more realistic cases, a detailed treatment of non-equilibrium  radiative cooling in the clouds (e.g., Melioli \& de Gouveia Dal Pino 2004; Melioli, de Gouveia Dal Pino \& Raga 2005) is required, particularly in the late stages of the core formation. The effects of  radiative cooling will be also considered in these forthcoming studies. 


\acknowledgments
MRML  acknowledges support from the Brazilian Agency CNPq (grants no. 140110/2008-9 and SWE 202114/2010-4),  EMGDP from the Brazilian agency FAPESP (grant no. 2006/50654-3) and CNPq  (grant no. 300083/94-7), and RSL from FAPESP (2007/04551-0). MRML also acknowledges the kind hospitality of A. Lazarian and his group during her three-month visit to the Astronomy Department of the University of Wisconsin. AL acknowledges NSF grant AST-1212096, Vilas Associate Award and the NSF Center for Magnetic Self-Organization in Laboratory and Astrophysical Plasmas. In addition, this study benefited from AL stay in the Universities of Cologne and Bochum enabled by the Humboldt Award as well as his stay at the International Institute of Physics (Natal, Brazil). The numerical simulations of this work were partially performed with the super cluster of the Laboratory of Astroinformatics (IAG/USP, NAT/Unicsul), whose purchase was made possible by the Brazilian Agency FAPESP (grant 2009/54006-4). The authors are also in debt to Grzegorz Kowal for his fruitful suggestions and constant advising regarding the employment of his code for the performance of the numerical simulations here presented. \textbf{The authors also acknowledge useful suggestions from an anonymous referee and very stimulating discussions with C. McKee, E. Falgarone and R. Banerjee.}


\appendix
\section{Theoretical justification of turbulent reconnection diffusion numerical studies on molecular clouds}
\label{theory}

\begin{figure}[!htb]
 \begin{center}
  \includegraphics[width=0.25\linewidth]{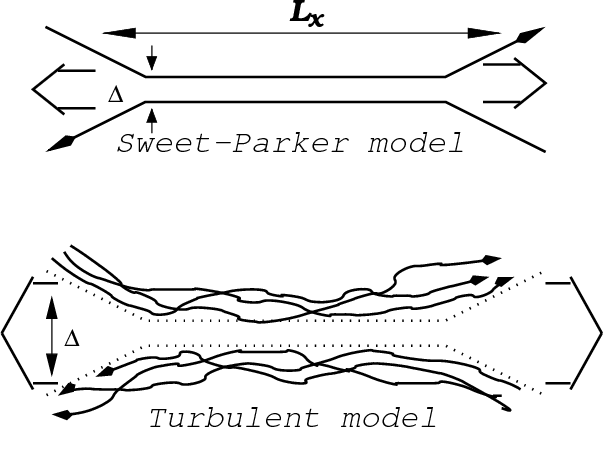}\\
\includegraphics[width=0.55\linewidth]{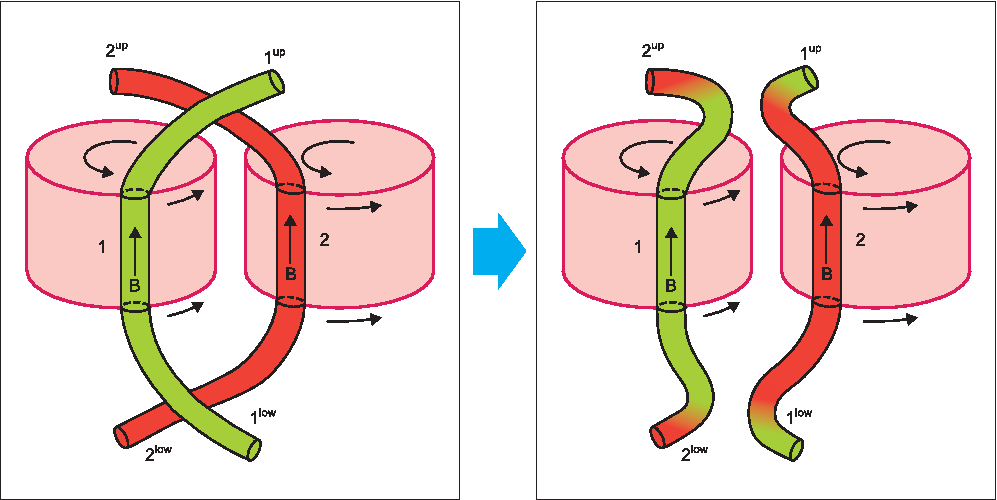}
\caption{Upper panel: Sweet-Parker reconnection scheme. Middle panel, schematic representation of LV99 model of fast magnetic reconnection. The LV99 reconnection rate does not depend on the micro-physics, but is determined by the turbulence properties (from Lazarian et al. 2004). Bottom panel: schematic illustration of the reconnection diffusion (RD) process. Two interacting turbulent eddies have their  magnetic flux tubes undergoing reconnection. Clearly, there is  exchange of parts of their flux tubes which naturally lead to magnetic flux diffusion in the  MHD flow (from Lazarian 2011).}
\label{fig:recon1}
 \end{center}
\end{figure}

The magnetic diffusion mechanism that we address here is related to MHD turbulence.

One should keep in mind that numerical simulations cannot exactly reproduce the reality of flows in molecular clouds (or any other astrophysical environment) which  are characterized by very high Reynolds numbers $Re=VL/\nu$, where $V$ and {\bf $L$ are the velocity and} the {\bf scale of} the flow, $\nu$ {\bf is the} viscosity. Equally  unaccessible are the high Lundquist numbers $S=L V_A/\eta$, where $V_A$ and $\eta$ are Alfv\'en velocity and Ohmic resistivity, respectively. The latter fact is of particular importance for our present numerical study that focuses on the diffusion of magnetic fields.

If magnetic field flux tubes do not reconnect they act as elastic bands which, though intersecting, create a network of magnetic knots that prevent magnetic field diffusion. Therefore, the issue of whether magnetic reconnection in turbulent fluids is fast or slow is of utmost importance for justifying the applicability of our numerical results obtained with small $S$ to the high $S$ molecular clouds.

The rates of reconnection in turbulent fluids were addressed in LV99 study. Figure \ref{fig:recon1} (middle panel) illustrates the process of fast reconnection that takes place {\bf in the presence of turbulence.} Unlike {\bf the Sweet-Parker (S-P) model} (upper panel) where {\bf the rate of reconnection is} limited {\bf by the} rate of plasma evacuation from the {\bf thin slot $\Delta$ determined by Ohmic} resistivity, {\bf the} outflow in the LV99 model {\bf is limited} only {\bf by the magnetic field wandering} induced by turbulence. Thus, the prediction of LV99 model is {\bf that the reconnection} is {\bf independent of} magnetic {\bf resistivity}. As most of the field wandering is due to the large scale eddies, LV99 model also predicts that the reconnection rates will not be sensitive to the extent that the actual turbulent cascade is resolved\footnote{The extent of the cascade depends on the $Re$ number of the flow, which is also much smaller than the real world molecular clouds. Fortunately, this is not an  important issue {\bf within the LV99 model of reconnection.}}.

In more quantitative terms, in the S-P model of magnetic reconnection, the {\bf velocity at which  two converging magnetic fluxes} of opposite polarity {\bf reconnect is given by $v_{rec} \approx v_A S^{-1/2}$. Because $S$ is large for Ohmic} resistivity values {\bf (e.g., for the ISM, $S \sim 10^{16}$), the S-P reconnection is} extremely {\bf slow}. In the LV99 model, on the other hand, it has been shown that turbulence is {\bf a universal trigger of fast reconnection.}  As a result, the process illustrated by Figure \ref{fig:recon1} (bottom panel) takes place and magnetic flux tubes and matter diffuse through turbulent eddies with velocities determined only by the properties of turbulence. Note, that it is proven in LV99 that within a turbulent cascade the reconnection happens within an eddy turnover time. This is an important justification of the relevance of the process in Figure \ref{fig:recon1} (bottom panel) to the actual high-$S$ reconnection in molecular clouds.

One can view our numerical simulations performed with relatively low $Re$ and $S$ numbers as large turbulent eddy simulations, where we resolve essentially the large scales. Although the correspondence with real molecular clouds should be provided also considering appropriate sub-grid physics, the key point  of the LV99 theory is that it predicts {\bf that the reconnection} rate {\bf does not depend on the} sub-grid physics and the important turbulent motions are represented by the large scale part of the turbulent cascade. This has been successfully tested with earlier much higher resolution numerical study (Kowal et al. 2009, 2012).  
 
Similarly, the correspondence of our present numerical studies with  the observed high-$S$ molecular clouds supports the concept of turbulent reconnection diffusion and provides indirect evidence for the LV99 process which is the core of the RD. In particular, we have shown that the effective RD coefficient is much larger than the numerical diffusion (see Section 4.2), so that the magnetic flux transport we detect, particularly in trans-Alfvénic clouds, is dominated  by RD rather than by numerical diffusion. 

\appendix
\section{Supplementary Figures}

\begin{figure}[!htb]
\centering
	\includegraphics[width=0.8\linewidth]{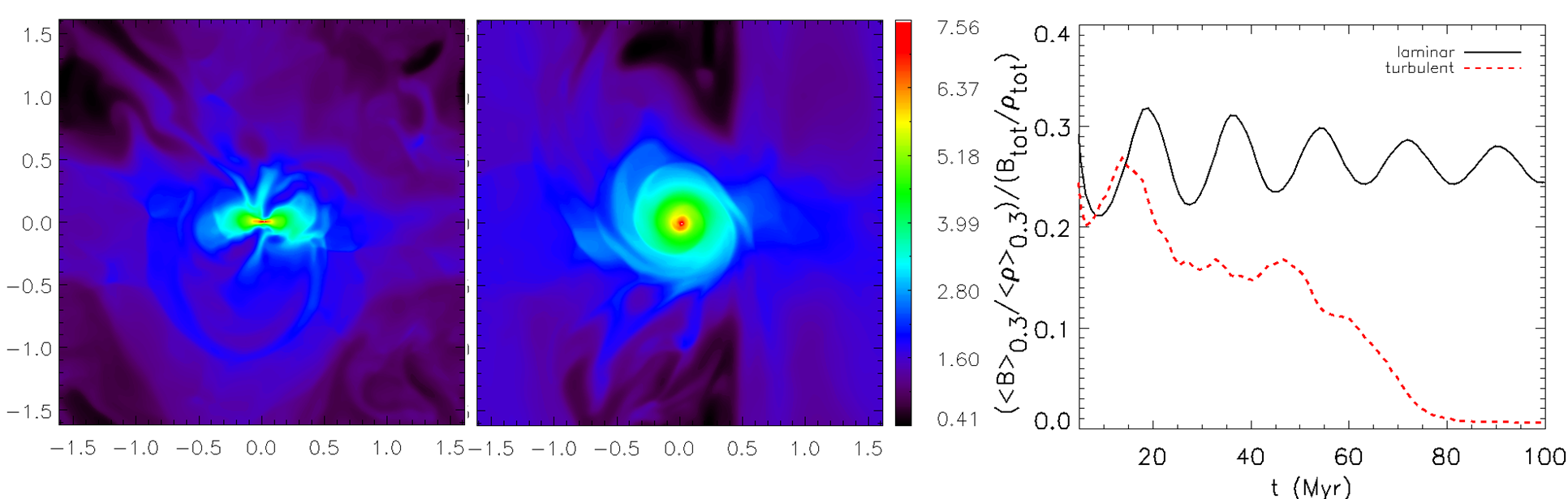}
  \caption{Model N2b-off. This model has the same initial conditions as in Model N2b, except that here turbulence was turned-off around 50 Myr. Left and middle panels show edge-on and face-on views, respectively, of the logarithmic density map of the central slice of the cloud at $t \sim 80$ Myr. The right panel shows the temporal evolution of the average magnetic field-to-density ratio at the cloud core region of radius $r_c= 0.3$ pc normalized by the average value over the entire cloud, $(\langle B \rangle_{0.3} / \langle \rho \rangle_{0.3})/(\bar{B}/\bar{\rho})$, for the turbulent (red-dashed lines) and the laminar (black continuous lines) cases (see the text for more details).
\textbf{We note that although the scale in the density diagrams indicate a  maximum density of $10^{7.56}$ cm$^{-3}$, this value is actually reached only within a few cells (2 to 3). The resulting density profile within the core region (with a radius $r<0.1$ pc) is so  sharp that the Jeans mass  required to its collapse is not achieved.}
 }
\label{fig:N2b_off}
\end{figure}

\begin{figure}[!htb]
  \centering	
    	\includegraphics[width=0.98\linewidth]{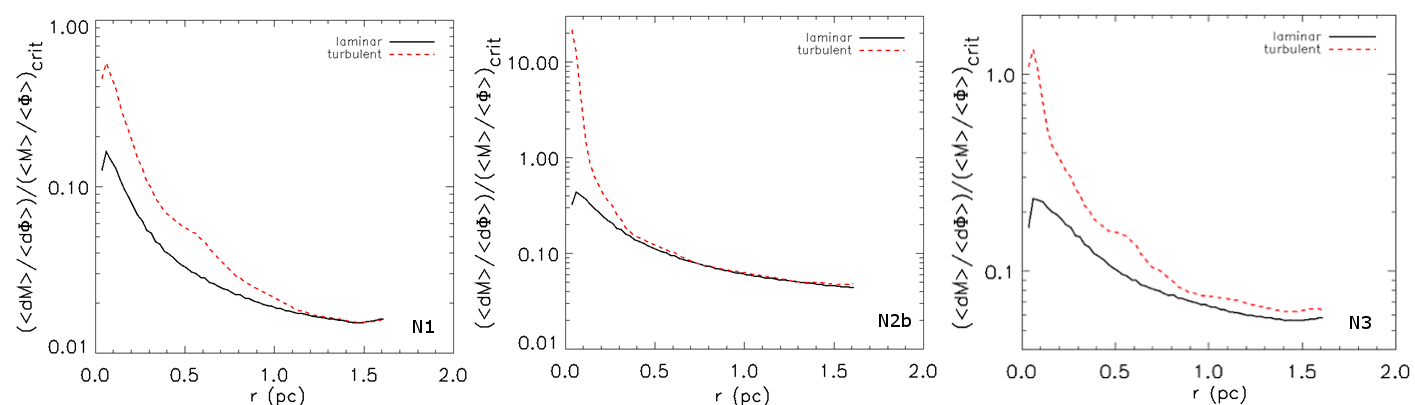}
  \caption{Radial profiles at $t= 100$ Myr for the 
magnetic flux-to-mass ratio normalized by the critical value $(dM/d\Phi)/(M/\Phi)_{crit}$. Left panel: model N1 ($n_0 = 10$ cm$^{-3}$, and $M_{\star}=61.1$M$_{\odot}$); center panel: model N2b ($n_0 = 90$ cm$^{-3}$, and $M_{pot}=40.7$M$_{\odot}$); right panel: model N3 ($ n_0 = 100$ cm$^{-3}$, and $M_{pot}=27.1$M$_{\odot}$). Red-dashed lines are for turbulent models and black continuous lines are for the laminar models. This differential ratio is clearly larger in the inner core regions than in the outer envelope of the clouds.}
  \label{fig:dMtodPhi}
\end{figure}

\end{document}